\documentclass[letterpaper]{article} % DO NOT CHANGE THIS
\usepackage{aaai2026}  % DO NOT CHANGE THIS
\usepackage{times}  % DO NOT CHANGE THIS
\usepackage{helvet}  % DO NOT CHANGE THIS
\usepackage{courier}  % DO NOT CHANGE THIS
\usepackage[hyphens]{url}  % DO NOT CHANGE THIS
\usepackage{graphicx} % DO NOT CHANGE THIS
\usepackage{natbib}  % DO NOT CHANGE THIS AND DO NOT ADD ANY OPTIONS TO IT
\usepackage{caption} % DO NOT CHANGE THIS AND DO NOT ADD ANY OPTIONS TO IT
\usepackage{algorithm}
\usepackage{algorithmic}

\usepackage{tabularx} 
\usepackage{booktabs} 
\usepackage{array}

\usepackage{newfloat}
\usepackage{listings}
\DeclareCaptionStyle{ruled}{labelfont=normalfont,labelsep=colon,strut=off} % DO NOT CHANGE THIS
\floatstyle{ruled}
\newfloat{listing}{tb}{lst}{}
\floatname{listing}{Listing}
\title{Abstracted Away: Resisting Alienation and Ungrounded Abstraction in AI Research Communities}
\author{
    Vyoma Raman\equalcontrib\textsuperscript{\rm 1},
    Isabel O. Gallegos\equalcontrib\textsuperscript{\rm 2},
    Neha Srivathsa\equalcontrib\textsuperscript{\rm 3}
}
\affiliations{
    \textsuperscript{\rm 1}Department of Information Science, Cornell University, USA \\
    \textsuperscript{\rm 2}Department of Computer Science and Stanford Law School, Stanford University, USA \\
    \textsuperscript{\rm 3}Department of Computer Science, Stanford University, USA \\
    vyoma@infosci.cornell.edu, iogalle@stanford.edu, nehasriv@stanford.edu
}

\begin{document}

\maketitle

\begin{abstract}

Logics of abstraction in computational AI research often push important forms of knowledge and reflection aside: dominant standards of legitimacy separate from lived experience of harm; the goals of work misalign with the practices that operationalize them; and career demands crowd out critical reflection. Even as prior academic and community-oriented efforts have sought to recontextualize and challenge common practices, exposure to sociotechnical harms and epistemic injustice persists. As three early-career critical AI researchers, we experienced this as alienation: feeling like outsiders in our research communities. This alienation has involved having some aspects of our backgrounds overlooked and others tokenized. We argue our alienation occurred through mechanisms that mirror abstraction by creating distance from relevant material realities. Beyond abstraction's role in computational AI research as a foundational practice structuring complex computational tasks, we have encountered it as a social norm in computational research spaces, illustrated through an autoethnographic inquiry into our alienation. We narrate three vignettes describing how we encountered and resisted alienation in our research communities. By analyzing themes across these accounts, we construct an interpretive framework of alienation categorizing its preconditions, mechanisms, and harms. Finally, we identify affect abstraction, one of the mechanisms of alienation we describe, as a high-leverage mechanism that is resistible by staying attuned to our affective responses, and collective action as a way to reduce risk and isolation when engaging in resistance. To assist others with similar reflection, we present our framework as a hermeneutic resource. Critical self-reflection and meaning-making are necessary steps toward challenging exclusionary disciplinary norms and cultivating more inclusive forms of AI research.

\end{abstract}


\section{Introduction}

In many introductory computer science (CS) classes, students learn to write their first functions. They are taught to name a procedure, specify its inputs and outputs, and --- crucially --- to set aside what happens inside other functions it calls. For beginners, this is often presented as a practical convenience, but embedded in this simple practice is a consequential idea. In CS, abstraction is the systematic practice of managing exposure to information about computational processes~\cite{Colburn2007Abstraction}. This is useful and necessary for tasks like writing functions, designing programs, and learning to model patterns in data. More broadly, abstraction reflects values core to computing, such as using modular design, structuring experiments, and specifying constraints of potential interventions. In this work, we define abstraction as both \textbf{elimination}, the omission of unnecessary details, and \textbf{essentialization}, the emphasis of important details, to model a phenomenon \textit{in service of a specified goal}. Both acts involve making some information more visible and other information less so, a framing highlighting that abstraction is inherently laden with value judgments regarding what kinds of details are important or unimportant for a given purpose. As these judgments are operationalized across systems and practices, they can become invisible and shielded from questioning~\cite{bowker2000sorting}.

Our investigation of the mechanisms of abstraction is an attempt to understand our shared experiences of \textbf{alienation} in computational AI research spaces. We define alienation as the implicit and explicit dissonance that comes with feeling out of place.\footnote{We intentionally select this terminology to invoke alienation from one's labor~\cite{Marx1844EstrangedLabourTucker}. Although our definition differs, both forms of alienation occur when people become estranged from the systems and environments where they work.} It is an emotional response to a perceived absence of \textbf{intellectual safety} in academic contexts, which includes having one's ideas and research valued and trusting others' intentions. Our observations of sociotechnical harm first triggered this alienation, when we experienced grief and indignation at witnessing the violences caused by AI systems, especially toward minoritized communities.\footnote{Drawing on multiple scholarly traditions~\cite{Dotson2011TrackingEpistemicViolence, farmer2004pathologies, galtung1969violence, galtung1990CulturalViolence, nixon2011slow, scheperHughesBourgois2004MakingSense, spivak1988subaltern}, we use the term ``violence'' to include physical, structural, cultural, epistemic, and other types of injustice and harm caused by sociotechnical systems, institutions, and asymmetries of power, while recognizing that these violences are not equivalent.} We also noticed that, in some cases, those most vulnerable to harm from AI systems shared some of our personal characteristics, which seemed obvious to us but unapparent to our colleagues. When challenging the design decisions causing these harms, we were subjected to various forms of \textbf{epistemic injustice}, which cast doubt on whether we belonged in computational AI research spaces. As put forth by \citet{Fricker2007EpistemicInjustice}, epistemic injustice delegitimizes people in their role as knowers, occuring when social power and prejudice distort how knowledge is produced, shared, or understood.

These experiences led us to search beyond computational AI to other disciplines, including law, disability justice, education studies, and health policy.\footnote{Our explorations beyond CS were made necessary by early challenges with identifying mentors to guide us on working on AI while staying aligned with our values. Interdisciplinary researchers face numerous challenges thriving within academia~\cite{Berkes2024SlowConvergence, Makinen2025Interdisciplinary, zheng2025interdisciplinaryphdsfacebarriers}.} In these fields, particularly their critical branches, we encountered alternate traditions that sensitized us to how judgments about which details matter can have significant real-world consequences. Later, we attributed the alienation we had experienced to the misalignment we perceived between which details were simplified across settings and the communities those simplifications were meant to serve. We do not reject abstraction as a whole; rather, we highlight the tensions that arise when an impulse to engage in abstraction is pursued without grounding in purpose. We use \textbf{grounded abstraction} to refer to practices that are intentional about which details are eliminated and preserved, considering the context of the abstraction. In contrast, \textbf{ungrounded abstraction} refers to practices that strip away details critical to the phenomenon being modeled, rendering the abstraction ineffective for the stated goal or creating material risks and disproportionate disadvantage in the stated context. We provide these and other key definitions in Table~\ref{table:definitions}, which we contextualize in Section~\ref{sec:framework}.

Our holistic experiences working on computational AI research are critical to our inquiry: they have enabled us to recognize signals of alienation and name the mechanisms that drove it. At the time of writing, we are graduate students at private universities in the U.S., trained in CS and working on interdisciplinary, critical machine learning and AI research. To synthesize our observations, we employ an autoethnographic method. The vignettes we present, narrated in a collective voice, are evidentiary material that we analyze to identify how ungrounded abstraction has alienated us. In the style of \citet{star2007enacting}, we blend the personal and analytic to formalize how ungrounded abstraction has driven our alienation in computational AI research spaces. 

\subsection{Roadmap and Contributions}
In the following sections, we narrate our experiences navigating AI research cultures and analyze these to identify how alienation manifests in our experiences and how we have resisted it. In Section~\ref{sec:approach}, we contextualize our autoethnographic method within a larger tradition of reflexive research practice. In Section~\ref{sec:v-clean}, we recount our first vignette: an encounter with the ethically fraught yet often unquestioned practice of data cleaning in an undergraduate class. In Section~\ref{sec:v-cv}, our second vignette, we depict experiences from the beginning of our graduate programs as we attempted to question assumptions and values underlying how computational AI research should interact with the world. In Section~\ref{sec:v-safety}, our third vignette, we highlight our experience engaging in an academic community where we found intellectual safety. While these vignettes touch on substantive issues including ethics education, discarding marginalized data points, human rights impacts of sociotechnical systems, and surveillance, we do not seek to intervene in each of them specifically. Rather, in Section~\ref{sec:framework}, we juxtapose the thematic similarities and differences in our experiences across these vignettes, creating a framework that characterizes how our alienation has operated. Specifically, we argue that \textbf{our alienation occurred through mechanisms that mimic the logic of abstraction by creating distance from relevant material realities.} Finally, in Section~\ref{sec:discussion}, we derive concrete interventions and broader strategies of resistance from our vignettes, our framework, and prior literature.

We ground our work in scholarship on reflexive methodologies (\S~\ref{sec:approach}) and prior efforts to synthesize structural issues in AI research (\S\S~\ref{sec:alien-lit}-\ref{sec:alien-abs-connect}). We also review prior, formally documented efforts to resist alienation within computational AI research spaces (\S~\ref{sec:discussion}), while acknowledging that much of this work lies in lived practice. This landscape shapes the contributions we claim:

\begin{enumerate}
    \item We describe our experiences of alienation and of finding solidarity in three vignettes (\S\S~\ref{sec:v-clean}-\ref{sec:v-safety}). These are evidentiary data and also acts of testimony.
    \item We present an interpretive framework of our alienation, connecting the preconditions, mechanisms, and outcomes of ungrounded abstraction (\S~\ref{sec:framework}). By highlighting structure that may be less visible in isolated accounts, we offer an interpretation of our alienation and provide a resource to help others engage in their own sense-making. We call back to prior literature in Sections \ref{sec:alien-lit}-\ref{sec:alien-abs-connect} that relate to the specific phenomena that triggered our alienation.
    \item We identify collective action and affective attunement as strategies to resist alienation and ungrounded abstraction. We frame this project itself as an intervention through the design of our methodology and output (\S~\ref{sec:discussion}).
\end{enumerate}

\begin{table*}[!t]
\centering
\small
\renewcommand{\arraystretch}{1} 
\begin{tabularx}{1.0\linewidth}{l X}
% \begin{tabularx}{\columnwidth}{l X}
\toprule
\textbf{Term} & \textbf{Definition}
\\
\midrule

Abstraction % notation
&  The superset of \textit{elimination}, the omission of detail, and \textit{essentialization}, the emphasis of detail.
 % definition
\\

Alienation % notation
&  The emotive experience of feeling out of place due to a lack of intellectual safety.
\\

Intellectual Safety
& The set of epistemic and relational conditions required to participate meaningfully in scholarship.
\\

Ungroundedness % notation
& The decoupling from real-world context and lived experience without justification, rendering something useless for the stated goal or generating material risks and disproportionate disadvantage in the stated context. % definition
\\

Metaeugenics % notation
& A set of logics that impact how nonconforming bodies are understood, treated, and regulated into compliance. % definition
\\

Legitimacy % notation
& The quality of those whom the field recognizes as a ``real'' AI researcher and the kinds of work it values. % definition
\\

Busyness % notation
& The demand of constant productivity and speed. % definition
\\

Testimony abstraction % notation
& The stripping away of one's legitimacy to contribute to knowledge-making. % definition
\\

Purpose abstraction % notation
&  The misalignment of the goal of pursuing a research project with how it is operationalized or evaluated. % definition
\\

Position abstraction % notation
&  The flattening of social relations, power asymmetries, and differences in lived experience in data and AI systems. % definition
\\

Affect abstraction % notation
&  The emotional distancing of a researcher from their work to better tolerate the harms they witness. % definition
\\

Epistemic injustice % notation
&  The undermining of someone in their capacity as a knower. % definition
\\

Sociotechnical harm % notation
&  Adverse effects toward affected communities in the social contexts in which AI systems are embedded. % definition
\\

Affective attunement % notation
&  The practice of intentionally noticing and responding to one's emotions as signals of real harm. % definition
\\
\bottomrule
\end{tabularx}
\caption{Concepts that have structured our analysis of alienation in computational AI research spaces.}
\label{table:definitions}
\end{table*}

\section{Approach}
\label{sec:approach}

Building on the traditions of feminist standpoint theory and critique of science, we take our experiences as an analytic resource. Feminist standpoint theory recognizes that marginalized social positions have unique perspectives on power and knowledge~\cite{anderson1995feminist, collins1989social, harding1992after, harding1992rethinking}. Situated knowledges make it possible to examine power relations~\cite{haraway1988situated, LugonesSpelman1983}. The resulting insights can be applied to critique scientific knowledge \cite{adam1993gendered, adam2000deleting, Code1991, DastonGalison2007}. Recognizing injustice as contestable makes it possible to forge alternative futures~\citep{freire1970pedagogy}.

Computational AI research cultures have long lacked intentional reflexive practice, but there have been a variety of attempts to close this gap. In 1997, \citet{agre1997critical} called for a \textbf{critical technical practice} that interrogates the premises and practices of AI technologies. Agre and others have autoethnographically recounted and analyzed their experiences to provide insight on the barriers and pathways to recontextualize AI and CS research \cite{agre1997critical, hofmann2020, khan2025whole, russo2024bridging, suchman2025ai, ymous2020terrified}. Other efforts have included applying critical theories to address existing practices \cite{hampton2021black, hanna2020towards, keyes2019misgendering, mohamed2020decolonial, shew2023technoableism}, theorizing how technical practitioners develop critical perspectives \cite{malik2022critical}, and encouraging reflection in research outputs \cite[e.g.,][]{beygelzimer2021introducing, naacl2022_responsible_nlp_blog, olteanu2023rai_impact_blog}. However, some analyses of these reflective practices suggest that authors insufficiently engage with potential impacts of their work, responsibility for harm, and how their positionality influences research outputs \cite{Liu2022Examining, Schroeder2025Disclosure}. Human-computer interaction research has also increasingly prioritized reflexivity in methods design and analysis \cite[e.g.,][]{Cambo2022Model, liang2021embracing}.

Our work builds on these efforts by reflexively examining our experiences in computational AI research communities, and producing an artifact that assists others in doing the same. Toward the former, we adopt an autoethnographic method because it explicitly centers groundedness and the specificity of individual experience. By centering situated lived experience, autoethnography operationalizes feminist standpoint epistemology and enshrines affect, shifting away from a ``view from nowhere''  whose emphasis on neutrality and detachment makes abstraction seem invisible and therefore inevitable~\cite{haraway1988situated, Nagel1986ViewFromNowhere}. As researchers immersed in computational AI research spaces, we can recognize the alienating effects of ungrounded abstraction on ourselves and analyze how these impacts occur.

\subsection{Method}

In autoethnography, researchers analyze their situated experience for insight into the conditions impacting them~\citep{ellis2011autoethnography, EllisBochner2000, ngunjiri2010living, StahlkeWall2016}, including in relation to technology~\citep{bala2023towards, kaltenhauser2024playing}. While autoethnography is often understood to emphasize emotive experiences~\cite{EllisBochner2000}, some interpretations prioritize systematic analysis of researchers' personal data~\cite{Anderson2006}. In our attempt, we identify similarities among our collective lived experiences and synthesize them to assist others with the same. We employ collaborative autoethnography, which performs self-interrogation within a group~\citep{bundy2023all, chang2016collaborative}. Some forms of collaborative autoethnography focus on individual reflective writing~\citep{ngunjiri2010living} and others emphasize group discussions~\citep{noel2023collective}; we used both approaches.

Our data collection process unfolded over six months, during which we recorded 16 hours of discussion focused on our experiences in computational AI research spaces and our interpretations of them. We summarized these conversations in detailed written notes and engaged in an iterative, inductive thematic analysis. 
This follows collaborative autoethnographic approaches~\citep{kafar2014autoethnography, nel2018relational, ngunjiri2010living, rutter2023its}. Each author independently conducted an open coding pass of the written notes to identify themes and produced short analytical memos synthesizing them in different ways. Following this, we had a series of discussions to reach consensus on the higher-level thematic groupings in the memos. From this analysis, we assembled three vignettes of our encounters with alienation that maximized coverage of the most salient themes. 
By articulating experiences that previously felt isolating and recognizing their shared structure, we lessened the alienation we had each carried. To reflect this experience, we describe our findings through vignettes narrated in a collective voice, representing multiple authors' accounts while remaining grounded in specific moments.

\section{Vignette: We Clean Ourselves from Data} \label{sec:v-clean}

We had a formative encounter with the dynamics of alienation and abstraction in an introductory data science course. As part of an assignment, we walked through data cleaning procedures to use medical records to model insurance risk scores. One step, presented as routine, required us to remove data points flagged as anomalous --- people whose data features deviated too far from the statistical norm. These cases were labeled noise. Excluding them, we were told, would clarify the ``real'' signal that the model was supposed to learn about the overall population, and the assignment stated that including them would make the model worse. In practice, this meant discarding people whose lived realities and medical histories did not align with the majority, reinforcing whose lives were worthy of prediction and whose were not.

It did not escape our notice that we were part of the groups being removed. We occupied the statistical tails that the assignment framed as expendable: people whose biomarkers fell outside general standards and whose identities were ``Other.'' The assignment forced us to enact an eliminative logic to filter out people like us. Rather than neutral preprocessing, we were being taught to routinely operationalize a logic of data ``cleaning'' that disciplined deviance.

A discussion question prompted us to reflect on the people represented in the data being dropped, recognizing that their removal could lead to worse model performance and downstream adverse effects for the groups they represented. But by proceeding with the ``cleaned'' dataset anyway, the assignment indicated that this should not alter the workflow. The underlying lesson: we were an unfortunate but inevitable tradeoff, since the model's performance for people like us ultimately did not matter. The brief interlude to discuss an ethical issue was just that --- an interlude --- sufficient to satisfy the instructional goal of teaching ethics while rendering followup changes to the approach unnecessary.

The stated purpose of this assignment was to teach trainee data scientists the best practices for effective predictive modeling, yet it offered no guidance on making principled tradeoffs between representing a diversity of people and isolating a ``clean'' pattern. By avoiding discussion of how to balance these competing aims, the assignment reduced a complex judgment to a simple rule in an act of ethics-washing. This left us with little intuition on how to apply the lesson to a realistic scenario. Our primary learning was that grounded reasoning could be acknowledged and then dismissed, and that contextual judgment was unnecessary.

Unable to contest a pre-programmed autograder, we had no choice but to rationalize this approach and suppress our discomfort. Data cleaning had been introduced to us as a necessary process to help models perform well on a desired population, and as students, we assumed that our unease reflected our inexperience rather than a substantive issue with the methodology. We trusted our professors to have considered perspectives like ours when designing the curriculum.

Later assignments in the course reinforced the learning from this one: the final project involved a leaderboard where credit was distributed according to standard performance metrics on a hidden dataset and where any approach was allowable. To succeed, we had to disregard potential downstream effects our methods might have on the individuals represented in the data. Under pressure to complete heavy projects in a tight timeline, it was surprisingly easy to file away our concern to revisit later. Thus, by decoupling substantive ethical reflection from the criteria judging our work, our professors abstracted away who the model represented and how. This left a decontextualized notion of model performance, not human impact, as the definition of success.

When the busyness of classes slowed and we regained our capacity, the marginalization embedded in the practices we had learned became clear. We had been taught to accept the exclusion of people like us as a ``reasonable'' sacrifice for the greater good, internalizing that our existence in the training data was incompatible with a useful model. This interpretation positioned our exclusion as computationally justified, a logic that extended far beyond us. No doubt similar justifications had been used in real-world contexts. The groups most likely to be removed from such datasets during data cleaning steps (such as disabled people, queer people, and racial and ethnic minorities) are also those most under-served by and vulnerable to many social systems. That is, the individuals abstracted away from the relevant AI systems would face greater material risk from any errors than the ones left in. By ignoring the impacts of flawed algorithms on people removed from the training data and adopting an approach that increased the likelihood of disproportionate predictive errors affecting those same groups, our introductory assignment implied that those most likely to face harms from a system need not be meaningfully represented within it.

\section{Vignette: Conflicting Visions for Computer Vision} \label{sec:v-cv}

In the first month of our graduate degrees, two encounters shaped how we experienced our research environments. The first was a significant escalation of colonial and imperialist violence, which we encountered through social media, journalistic reporting, and conversations with those who were personally affected and organizing to protest it. Media coverage increasingly brought to light how AI technologies, including computer vision (CV) tools, were being used for military purposes. The second encounter was a paper that traced the connection between CV research and downstream surveillance applications~\citep{kalluri2025computer}. This work illuminated how CV research conducted in academic and other settings, including application-agnostic technologies or those built for beneficial and benign applications, can simultaneously be used for oppression. It sickened us that efforts to build AI for social good could be corrupted like this --- and that such dual uses have characterized the field since its beginnings~\citep{leslie1993cold, selinger2022amazon, waelen2024ethics}. Even as reports on such destructive applications became widespread, we observed that they were still overlooked in the CV research discussions around us, which emphasized potential benefits over associated risks. 

These threads began to overlap. Our peers drew attention to how surveillance technologies were already all around us, especially in areas of campus most frequented by minoritized students. We observed the embodied effects of these surveillance technologies: we changed our movement patterns, were hyper-aware of cameras and potential surveillance devices when traversing campus, and carried tension that we didn't before. Activists pointed to the use of CV in the ongoing imperialist escalation, heightening our awareness of how research enabled the devastating violence reported in the news. These observations raised unavoidable questions about the research that we were conducting and the ways that the research careers we had at one point envisioned for ourselves could enable oppressive regimes.

At the time, we were each working on advancing AI capabilities in different areas, hoping to leverage technology to improve lives. Facing the reality that many AI systems are implicated in violence, we questioned whether the research agendas that we were contributing toward would enable oppression~\citep{kalluri2025computer, stark2019facial, suchman2015situational}. We worried that working on research to improve AI would require complicity. These questions weighed on us heavily. We found ourselves bringing up these topics in conversation, hoping to understand our peers and mentors' mental models of impact and how they reconciled ethical dilemmas.

Among fellow graduate students, we saw curiosity about the social impacts of research, tempered by an emotional distance. We felt that these concerns were more central to our own day-to-day lives in academia than our peers'. Like others, we experienced a constant stream of research, teaching, and administrative tasks, and faced the speed of publishing cycles in AI. But despite pressures to gain credibility as early-career researchers, working on specific topics with specific people at the cost of decentering ethics from our research agendas was not a tradeoff we were willing to make.

When discussing our dissonances with more established researchers, we received guidance illustrating the constraints of dominant research agendas. We were encouraged to build ``ethical'' versions of AI technologies. We saw this as reflecting an underlying belief that developing such tools, as currently envisioned, was inevitable, and that ethical responsibility mainly consisted of incremental improvements. While aligning with these agendas would undoubtedly make certain career paths, such as CS academia, more attainable, we were concerned that this would require us to decenter the values that we hoped to advance through our research, which prompted our interest in graduate school in the first place. Although we acknowledged the care that this advice stemmed from, we were in search of guidance on how to resist the pressure to conform to prevailing research agendas without being perceived as naive and dismissed as scholars.

In considering the mission of developing more ethical versions of AI technologies, some researchers cautioned us against being positioned as ``merely'' the ``ethics person'' in our respective areas, warning of tokenization and minimization of our computational skills. These comments led us to question whether critical work in these spaces truly had the power to transform dominant computational agendas, or whether it would only be granted legitimacy when affirming existing research agendas. These reflections felt sobering.

We had hoped that these conversations would support our collective growth alongside our colleagues and help us refine our research goals. While we felt cared for, we left unsure how to proceed. The questions we were asking felt impermissible in many research spaces around us, and the responses we received often reflected assumptions and values about AI development that we disagreed with. These premises felt difficult to escape. We were left longing for a community to explore such questions with.

\section{Vignette: Intellectual Safety in Community} \label{sec:v-safety}

Upon feeling alienated from computational AI research communities, we began searching for alternate spaces where our concerns would be constructively centered as respectable research inquiry. We sought to critique computational advancements and envision liberatory computational futures in intellectually safe environments. We experienced some successes in finding community: the authors met each other and identified mentors who had had similar experiences. Some mentors introduced us to others, but eventually, despite our desire for more intellectual companionship, our networks of critical scholars in CS stopped growing.

During this period, we found a community of critical technology scholars based out of Stanford's Graduate School of Education.
There, we found peers grappling with similar topics from diverse disciplinary vantage points including, but not limited to, CS.
When a paper describing the eugenic ideologies underlying modern AI \cite{gebru2024tescreal} was released, one coauthor volunteered to host a reading group session within this community to discuss this paper. RSVPs, which included the other two coauthors, spanned scholars in CS, education, science and technology studies (STS), communication, and sociology. The ensuing event was lively: discussion topics ranged from dissecting the paper's argument to analyzing its discursive intervention. Due to the diverse disciplinary backgrounds in the group, each attendee brought a different perspective and area of expertise. Each person had to translate the methods, concepts, assumptions, and ideologies of their discipline for others, intentionally abstracting away some context to make their point legible.

After the event, the three authors walked to our workspaces together. As we did so, we mused about why this reading group had felt so positive, in contrast to the alienation that we persistently felt in computational AI research spaces. We each experienced an embodied sense of ease in the space, so distinct from the tension we often carried in CS spaces. We identified three notable differences. First, we noticed the pace of conversation was slowed; people were respectful of others' speaking time, left space to reflect, and engaged with what others had shared. Second, due to the interdisciplinarity of the group, participants did not take assumptions for granted, instead spending time to establish a common knowledge base. Third, it was not only acceptable but welcome to critique the values and assumptions of the work. Altogether, these norms created a markedly different intellectual experience and affirmed our desires for more.

Our conversation during this walk was the first time we had discussed our alienation as a group rather than in pairs. Identifying the parallels between our experiences helped us validate our feelings of alienation. Further, we became aware that frustrations, which previously felt like isolated experiences or coincidences, might be caused by structural, cultural, and institutional factors. We began to consider that, beyond merely coping with our alienation, we could examine and even actively resist the factors impacting it.

At the time, the three of us were at different transition points: some nearing the end of PhD rotations and deciding next steps, and some deciding which next degree programs would serve our intellectual needs. We sought to replicate the microcosm of our discussions on that day at a greater scale. We were motivated by the intuition that being part of a larger community that does careful and critical AI work would help us become the scholars we wanted to be and do the kind of research that would push AI in the directions we wanted. This conversation between the three of us, where we affirmed each of our respective goals, identified the structural factors driving our experiences, and considered how logics of eugenics may permeate AI systems and communities, was, in retrospect, an early inception of the work that forms this paper. 

\section{Analyzing Our Alienation}\label{sec:framework}

We examine the factors that created or averted our alienation in each vignette, finding them to follow similar patterns that we characterize in this section.

\subsection{Alienation in Context} \label{sec:alien-lit}

We experienced alienation as an emotional reaction to recurring dynamics experienced across coursework, meetings with collaborators, and other contexts. The contrast between our experiences of alienation and the lack thereof is depicted in the previous sections and expanded upon in Appendix \ref{app:themes}. These reveal underlying tensions that perpetuate alienation that have been extensively theorized by various literatures.

\subsubsection{Interdisciplinarity} \label{sec:alien-lit-interdisciplinary}  
Prior literature has explored the challenges of legibility in interdisciplinary research. Interdisciplinarity is celebrated for its ``essential tension'' of bringing together different fields for disruptive innovation~\citep{kuhn1977essential}. It requires translating between epistemologies and methodologies, intellectual values and cultures, and organizational and institutional traditions~\citep{bauer1990barriers, fahimi2024articulation}. Interdisciplinarity is harder still in the context of computational AI research,  which has particular tensions with humanistic and socially-oriented fields ~\citep{fahimi2024articulation, klein2025provocations}. These tensions include demands for formalization over open-ended interrogation; fast progress on well-defined problems over slower inquiry; and generalizability over contextual grounding~\cite{klein2025provocations, klumbyte2022critical}. Beyond creating mutual friction between fields, interdisciplinary work can seek to avoid challenges to dominant perspectives, leading to systematic marginalization of alternative approaches~\citep{klein2025provocations}.

\subsubsection{Institutional Logics in AI Ethics} \label{sec:alien-lit-inst} 

The struggles in our vignettes emerged when tensions that felt unavoidable to us clashed with institutional logics that deemed our concerns as unimportant. Prior literature has described how dominant AI ethics work is institutionally situated, influenced by a combination of corporate, academic, and political power \cite{bietti2020ethics, green2021contestation, metcalf2019owning, young2022confronting}. This impacts what is recognized as legitimate AI ethics work and who is able to conduct this work. Under institutional logics, AI ethics can become oriented toward legitimizing AI advancement rather than meaningfully influencing or constraining it \cite{green2021contestation}. 

Minoritized communities are systematically underrepresented in AI ethics, in the consideration of harms~\cite{birhane2022forgotten} and in their participation in shaping the field. AI researchers may define ethics in ways that strip context away~\cite{selbst2019fairness}, leading to obscured power differentials that impact minoritized communities. AI ethics work that is based in lived experience, often conducted by people from minoritized groups, is delegitimized, while other AI ethics work uses quantification as a strategy for legitimacy \cite{Widder2024Epistemic}. This privileging of computational ways of knowing reproduces epistemic exclusion even at the level of CS education \cite{raji2021you}.

\subsubsection{Legitimizing Situated Knowledges} \label{sec:alien-lit-positionality}

Tensions around which situated knowledges are legitimized or dismissed persist, even as many venues encourage the integration of positionality and reflexivity into research outputs~\cite{aies2026authorguide, facct2026authorguide}. Formal positionality statements have been found to insufficiently connect to research outputs, and they are still interpreted and implemented inconsistently~\cite{Schroeder2025Disclosure, Singh2025Exploring}. Empirical findings further complicate this: while these venues have diversified the research topics they publish, they continue to reproduce structural inequities in whose knowledge is represented~\citep{acuna2021are}. This suggests that researchers differ in their ability and desire to reflexively change their research practices in the ways that such measures seek to promote. As such, much work remains to legitimize diverse knowledges in AI research.

\subsection{From Alienation to Abstraction} \label{sec:alien-abs-connect}

The vignettes illustrate our experiences of alienation as not just a lack of care or a mismatch between our values and those of our research communities, but as a product of pressures we faced to conform to a research agenda that prioritized contributions to AI development over those to society, elevated potential benefits of technology over harms, and emphasized optimization on generalizable problems while leaving underlying normative commitments unstated or misspecified~\citep{Birhane2022Values, Laufer2023Optimization}. While we initially attempted to oblige, the stress of pretending that harm wasn't occurring took a toll on us that did not alleviate until we began to address our complicity. We were further expected to enact modes of research that assumed a view of technological progress centered on values like generalization, efficiency, benchmark performance, and novelty on large datasets~\citep{Birhane2022Values}, pursuing our work only within them, when what we wanted was to interrogate the harms inherent to this status quo and imagine alternate technological futures. Sharing these experiences with each other, we began to realize their structural components and carve out a space of intellectual safety without these constraints.

This pressure toward conformity resembles a regime that ~\citet{Williams2019Metaeugenics} characterizes as \textbf{metaeugenic}: containing ``cultural norms, ideals, values, and demands that warp and twist deviant bodies into conformity.'' Encapsulating the logics of sanitization, normativity, and compliance, metaeugenics is a thread that summarizes our experiences of alienation and has been established in the contexts of computational cultures and systematic oppression~\cite{Chan2025Predatory, gebru2024tescreal, Williams2025Disabling}. We apply metaeugenics to study how deviance in our computational AI research communities is disciplined into conformity.

Conformity requires stripping away one's ``aberrant'' characteristics, an act reminiscent of abstraction's process of omitting certain details. To understand how abstraction sustains a metaeugenic regime in computational AI research communities, we consider its normative nature. Much scholarship has questioned which and whose values are embedded in computational systems and how abstraction can obscure power asymmetries and harms \cite{Laufer2023Optimization, selbst2019fairness, Wang2025Identities}, reminiscent of earlier accounts of idealized formalism displacing embodied experience~\cite{husserl1970crisis}. In addition to influencing computing technologies, abstraction shapes CS by enabling researchers and developers to remain detached from the contexts where their outputs will be used, starting as early as introductory CS education \cite{Birhane2022Values, malazita2019infrastructures, peterson2023abstracted}. More broadly, philosophers of science have emphasized that values shape both scientific research and its societal consequences~\cite{douglas2000inductive, longino2020science}. Together, this work suggests that abstraction mediates values within computational research communities and in sociotechnical artifacts that impact the world. Even so, abstraction's potential for harm or benefit depends on how much it is grounded in context and goals. Our vignettes make these impacts visible in computational AI research by showing how it, in our experiences, shapes not only research outcomes but also participation, legitimacy, and belonging.

\subsection{Interpretive Framework of Alienation} \label{sec:framework-components}

\begin{figure*}
    \centering
    \includegraphics[width=0.8\linewidth]{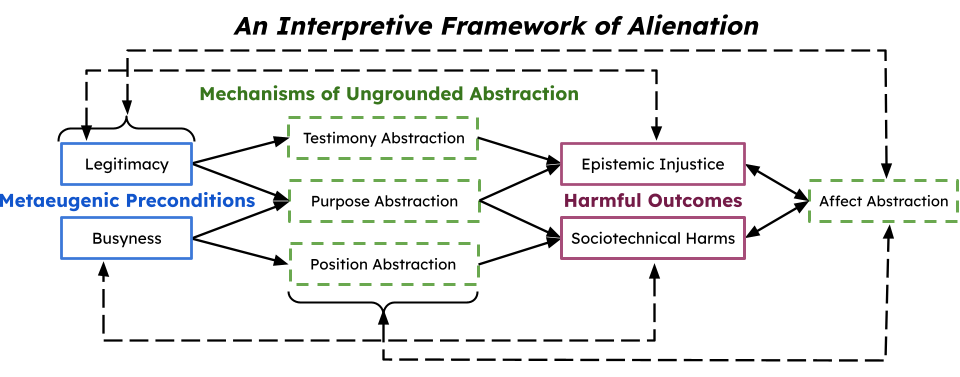}
    \caption{Interpretive framework of alienation, as present in our own experiences. We identify two metaeugenic preconditions that enable four mechanisms of ungrounded abstraction, which generate two categories of harmful outcomes. Arrows imply proposed causal relationships. Directed arrows indicate that the source node impacts the destination node, whereas bidirectional arrows indicate mutual influence between the two nodes. We highlight a pathway from preconditions to outcomes with solid arrows; dashed arrows represent how the components perpetuate each other.}
    \label{fig:framework}
\end{figure*}

We propose a framework (Figure \ref{fig:framework}) that links our experiences of alienation to patterns of abstraction, the conditions that sustain them, and the harms they produce. This framework reflects thematic similarities across our three vignettes that emerge from our coding of them (see Appendix \ref{app:themes}).

Our framework contains three types of components: (i) \textbf{metaeugenic preconditions} that shape research culture; (ii) \textbf{mechanisms of ungrounded abstraction} that operate within it; and (iii) \textbf{harmful outcomes} resulting from these mechanisms that affect researchers and society. We connect each component to our experiences with an alienating example from one of the first two vignettes and a counterexample from the third, listing the corresponding vignette with each example. We also briefly characterize the preconditions and mechanisms using prior literature.

\paragraph{Metaeugenic Preconditions}
Drawing on prior work (\S~\ref{sec:alien-abs-connect}), we identify metaeugenic preconditions as critical to shaping our experiences of computational AI research. We frequently saw two particular  norms enabling the abstracting mechanisms in our experiences:
\begin{itemize}
    \item \textbf{Legitimacy} governs who the field recognizes as a ``real'' AI researcher and what kinds of work it values. Research that emphasizes methodological novelty, disciplinary purity, and alignment with dominant research agendas is seen as legitimate, while work studying lived experience, interdisciplinary critique, or adverse consequences within a context is often secondary or out of scope.
    \item[] \textit{\underline{Prior literature}}: Different fields have distinct epistemologies and approaches that implicitly exclude certain ways of knowing~\citep[\S~\ref{sec:alien-lit-interdisciplinary},][]{bauer1990barriers, fahimi2024articulation}, and legitimacy can systematically foreclose alternate approaches~\citep{klein2025provocations, klumbyte2022critical}. The institutional logics of AI ethics (\S~\ref{sec:alien-lit-inst}) further legitimize harmful AI work rather than constraining it~\citep{green2021contestation}. We draw from these ideas that center power, where structural asymmetries determine whose research agendas are recognized as belonging to the field.
    \item[] \textit{\underline{Example (\S~\ref{sec:v-cv})}: Mentors warned us about the organizational distinction some labs make between ``technical'' researchers who advance AI capabilities and ``ethics'' researchers who study or mitigate their oppressive uses.}
    \item[] \textit{\underline{Counterexample (\S~\ref{sec:v-safety})}: Event participants translated between different disciplines, enabling broader engagement in critical discussions of AI.}
    \item \textbf{Busyness} demands constant productivity and speed. Grounded work, especially work that requires engagement with affected communities or ethical uncertainty, is often positioned as inefficient or unnecessary.
    \item[] \textit{\underline{Prior literature}}: Institutional logics encode capitalist and corporate structures that promote metricized, productivity-oriented ways of working~\citep{bietti2020ethics, green2021contestation, metcalf2019owning, Widder2024Epistemic} over slower, critical engagement~\citep{klumbyte2022critical}. Our notion of busyness connects this to the pace demanded by these pressures.
    \item[] \textit{\underline{Example (\S~\ref{sec:v-clean})}: The fast pace and heavy workload of our CS courses prevented us from engaging more deeply with the implications of the techniques we learned.}
    \item[] \textit{\underline{Counterexample (\S~\ref{sec:v-safety})}: Discussion proceeded at a pace that allowed participants to fully express their ideas and reflect before engaging, without fear of being left behind.}
\end{itemize}

\paragraph{Mechanisms of Ungrounded Abstraction}
We identify four types of abstraction that reinforce the metaeugenic regime: testimony abstraction, purpose abstraction, position abstraction, and affect abstraction.

\begin{itemize}
    \item \textbf{Testimony abstraction} concerns who is heard. Abstraction strips away one's legitimacy to contribute to knowledge-making in computational AI research spaces. Researchers' credibility is selectively amplified or diminished by framing their expertise or experience in ways that reify disciplinary boundaries. Critique is dismissed either because a researcher is deemed insufficiently ``technical'' or because their concerns are framed as naive, emotional, or inevitable. In these cases, critical scholars of computational AI research experience testimonial injustice~\cite{Fricker2007EpistemicInjustice}.
    \item[] \textit{\underline{Prior literature}}: Prior studies show that the work of researchers from minoritized communities is disproportionately delegitimized~\citep{birhane2022forgotten, raji2021you, Widder2024Epistemic}. \citet{Fricker2007EpistemicInjustice}'s account suggests that this is a structural reduction of those researchers' credibility.
    \item[] \textit{\underline{Example (\S~\ref{sec:v-cv})}: We worried about being perceived as naive when we questioned the necessity of building certain AI technologies, based on their violent uses.}
    \item[] \textit{\underline{Counterexample (\S~\ref{sec:v-safety})}: Every researcher added value to the discussion due to their distinct background.}

    \item \textbf{Purpose abstraction} concerns why an action is taken. Abstraction occurs when the reason for pursuing a project becomes misaligned with how it is operationalized or evaluated. Methods may drift away from the problem they claim to address, and may deviate in ways that actively oppose the original intent. Career incentives, prestige, or novelty may impede substantive engagement with the original goal. 
    \item[] \textit{\underline{Prior literature}}: The purpose of research has been found to diverge from its methods when translation across epistemologies and institutional cultures breaks down during interdisciplinary collaborations~\citep{bauer1990barriers, fahimi2024articulation, klumbyte2022critical}. AI ethics frameworks may be applied superficially to serve the advancement of the research they are meant to scrutinize~\citep{green2021contestation, Widder2024Epistemic}.
    \item[] \textit{\underline{Example (\S~\ref{sec:v-clean})}: In our coursework, the goal of training ethical computer scientists was diluted when the assignment chose to merely name ethical issues instead of modifying approaches to address them.}
    \item[] \textit{\underline{Counterexample (\S~\ref{sec:v-safety})}: The group engaged with the underlying assumptions and values of the paper as well as its other components; this intentionally holistic critical discussion made space for wide-ranging examination.}

    \item \textbf{Position abstraction} concerns who is affected. Abstraction removes social relations, power asymmetries, and differences in lived experience from data and sociotechnical systems. It may erase imbalanced power or assume uniformity and interchangeability between developers, users, and subjects of AI systems. Together, these obscure how marginalized groups may be disproportionately impacted in ways that exacerbate inequities.
    \item[] \textit{\underline{Prior literature}}: Previous interventions to encourage reflexivity and analysis of situated knowledge (\S~\ref{sec:alien-lit-positionality}) include positionality statements specified by publishing venues~\cite{facct2026authorguide, aies2026authorguide}. While these recognize that researcher positionality shapes research, it is difficult for an intervention that frames positionality as a disclosure problem to encourage active reflection on how one's position and power meaningfully shape the research process \cite{Liu2022Examining, Schroeder2025Disclosure, Singh2025Exploring}. Institutional logics compound this by limiting access to who can meaningfully shape the field~\citep{bietti2020ethics, birhane2022forgotten, green2021contestation, metcalf2019owning} and by removing the contextual specificity that makes disproportionate impacts visible~\citep{selbst2019fairness}.
    \item[] \textit{\underline{Example (\S~\ref{sec:v-clean})}: We filtered from the training data people whose positionalities both made them candidates for removal and would uniquely make them vulnerable to harms from the AI system.}
    \item[] \textit{\underline{Counterexample (\S~\ref{sec:v-safety})}: The discussion's chosen emphasis on eugenics foregrounded considerations of deviance and who is considered valuable.}

    \item \textbf{Affect abstraction} concerns how researchers feel about the effects of their works. It occurs when researchers emotionally distance themselves from their work to better tolerate the harms they witness without disrupting the prevailing metaeugenic norms. Each of the other types of abstraction enables this distancing by reframing adverse consequences as a computational necessity or acceptable trade-off. By enforcing emotional discipline, affect abstraction renders metaeugenic norms as natural and the harms produced by other types of abstraction as inevitable. In contrast, \textbf{affective attunement} involves remaining responsive to emotions as sources of epistemic and ethical insight; by making harm emotionally visible, attunement destabilizes the metaeugenic regime in ways that are difficult to dismiss. 
    \item[] \textit{\underline{Prior literature}}: Affect theory broadly explores embodied responses to one's experiences, including how social power operates within environments~\citep{clough2007introduction}. Reflexive practice in research taps into these and other experiences as sources of knowledge~\citep{hampton2021black, hanna2020towards, keyes2019misgendering, mohamed2020decolonial, shew2023technoableism}. Our conception of affect abstraction emphasizes the role of affect in identifying mechanisms of injustice~\citep{Lorde1981} and the harm that occurs in the absence of emotional attunement.
    \item[] \textit{\underline{Example (\S~\ref{sec:v-cv})}: When we learned about CV applications that threatened human rights, we were encouraged to suppress our discomfort with the potentially harmful uses of the technologies we worked on.}
    \item[] \textit{\underline{Counterexample (\S~\ref{sec:v-safety})}: Collectively articulating experiences of feeling out of place helped us identify the recurring emotional patterns shaping our alienation.}
\end{itemize}

\paragraph{Harmful Outcomes} 
We connect these preconditions and mechanisms to two established classes of harmful outcomes that have perpetuated different violences on researchers and communities impacted by sociotechnical systems.
\begin{itemize}
    \item \textbf{Epistemic injustice} causes researchers' knowledge and credibility to be unfairly dismissed \cite{Dotson2011TrackingEpistemicViolence, Fricker2007EpistemicInjustice}. This arises from testimony abstraction that challenges the legitimacy of critics, and from purpose abstraction that evaluates researchers with metrics decoupled from their intellectual contributions.
    \item[] \textit{\underline{Example (\S~\ref{sec:v-cv})}: We were cautious about being seen as the ``ethics people'' in our labs, which might dismiss our technical expertise and minimize our work.}
    \item[] \textit{\underline{Counterexample (\S~\ref{sec:v-safety})}: Our community of critical technology scholars took seriously our contributions as both technical and ethical experts.}
    \item \textbf{Sociotechnical harms} toward affected communities arise when AI systems produce adverse impacts in the social contexts in which they are used~\cite{10.1145/3600211.3604673}. These occur through position abstraction that prevents AI practitioners from adequately understanding the social relations surrounding AI systems, and through purpose abstraction that permits methods that may conflict with the purported social goal of a system.
    \item[] \textit{\underline{Example (\S~\ref{sec:v-cv})}: Militaries use computer vision weapons against populations subjected to colonization.} 
\end{itemize}

While this framework reflects how ungrounded abstraction perpetuated alienation in our own experiences, we emphasize that our goal is \textit{not} to offer a generalized theory. Rather, we offer this framework as a generative resource to assist others in making meaning of their own experiences. Given the shortcomings of some existing approaches to researcher reflexivity \cite{Schroeder2025Disclosure, Singh2025Exploring}, leveraging this framework may help deepen reflection. We further describe the relationships between the nodes in our framework in Appendix \ref{app:relations}, to further expand on their mutual interactions.

\section{Discussion}
\label{sec:discussion}

\paragraph{Ungrounded Abstraction Fuels Alienation.}
Our framework links moments of alienation from our research communities to the abstraction mechanisms through which they occurred. Each form of ungrounded abstraction eliminated or essentialized the contexts that grounded our work. As a result, our critiques lost salience. Much as notions of objectivity, neutrality, and the ``view from nowhere'' strip away the knower, each form of ungrounded abstraction removes one's accountability to situated experience. While our framework calls upon feminist standpoint epistemology, \textbf{our analysis of alienating experiences identifies a logic of abstraction within our computational AI research communities}. Ungrounded abstraction actively produces the conditions of alienation by removing testimony, purpose, position, and affect from everyday research practice. Thus, our alienation signaled when the presence of ungrounded abstraction pressured us to conform to a dominant research agenda.

Our critique is not unique to computational AI research cultures, but we argue that computational AI research amplifies the dynamics of alienation in distinctive ways. Particularly, AI enables the rapid operationalization of abstractions into widely deployed sociotechnical infrastructures \cite{selbst2019fairness}, and AI research cultures privilege mathematical and quantitative formalizations over other forms of knowing, which can make contestable assumptions seem objective or inevitable~\cite{adam1993gendered, agre1997critical}.

\paragraph{Proposed Interventions}
By connecting ungrounded abstraction to metaeugenic preconditions, our framework of alienation indicates that there exist interventions that can act on both. We suggest avenues for this below\footnote{Many factors affect the ability to change research practices. We draw on the framework and vignettes to propose interventions, so we are inherently limited by the scope of our method and do not claim our suggestions to be universal.} and invite others to interact with our framework by using their personal experiences to identify new practices.

The accessibility of interventions may vary across career stages, reflecting differing power and risk. More established researchers are positioned to reject metaeugenic ideals of busyness and legitimacy. They can resist busyness by eschewing rapid publishing cycles that incentivize the sacrifice of depth and contextual grounding. In their labs, departments, and relevant research associations, they can encourage the questioning of disciplinary premises. To broaden standards of legitimacy within the field, they can platform a range of perspectives at events and use their power as reviewers or editors to affirm the value of topics or methods left out of dominant scholarly discourse. To create protective conditions for early-career scholars to engage in resistance, more established scholars can mentor them on how to navigate institutional structures while encouraging them to question those structures and develop the skills needed to transform them. This could include creating transparency around how decisions influencing academic spaces are made and making visible the hidden curriculum of academia. 

Early-career researchers, especially students, are primarily responsible for executing projects. This empowers them to avoid ungrounded abstraction in everyday research decisions. For instance, they can resist position abstraction by being intentional about how they operationalize groups and identities within datasets; testimony abstraction by centering scholarship from minoritized researchers; purpose abstraction by assessing whether design decisions align with the goal of their research projects; and affect abstraction through deliberate affective attunement, a strategy we detail below.

Even as we believe in the power of all people to transform research norms, the labor of resistance is often distributed unevenly. Students are particularly at risk for making research decisions that conflict with the direction of superiors, while remaining susceptible to distress from engaging in ungrounded abstraction. We encourage researchers at all levels to reflect on their positionalities and the forms of resistance accessible to them. To manage these risks, we identify collective action and affective attunement as key strategies.

\paragraph{Resisting Alienation Through Collective Action.}
We identify collective action as essential for resisting mechanisms of alienation, especially given our experiences in Section~\ref{sec:v-safety}. Collective action can distribute the risks of deviating from norms. When early-career researchers exercise their power to challenge practices collectively, retaliation becomes more difficult. Approaches can include sharing experiences privately and publicly, coordinating pushback on harmful norms, sharing risk through joint statements, and creating visible spaces of collective critique, such as reading groups or workshops. Visible critical spaces can also serve as gathering points for organizing collective power. Communities in technology spaces are already engaged in important forms of collective action, including labor organizing efforts like the Tech Workers Coalition, identity-based communities like Queer in AI, data governance initiatives such as the Indigenous Data Sovereignty Network, academic collectives like the Stanford Critical AI Group, and geographically distributed events like Deep Learning Indaba.

In addition to critiquing harmful structures, transforming research practices requires envisioning and creating alternatives. This involves \textbf{futuring} and \textbf{prefiguring} desired research communities by enacting in the present the ethics and epistemic practices that guide the versions to be realized in the future~\cite{boggs1977revolutionary}. We participated in an iteration of this strategy (\S~\ref{sec:v-safety}) by creating a community grounded in interdisciplinary critique. Further, as we experienced with each other, community support can enable shared interpretation of alienating experiences, which may otherwise remain obscured by affect abstraction. Community can protect researchers by distributing risk and creating the conditions for affective attunement that may be less accessible in isolation.

\paragraph{Resisting Ungrounded Abstraction Through Affective Attunement.} 
Inspired by \citet{Lorde1981} and \citet{ahmed2004cultural}, and building on the humanities' ``affective turn'' that situates embodied reactions within systems of labor, technology, power, and control~\cite{clough2007introduction}, we call for \textbf{affective attunement}, the practice of intentionally cataloging one's emotions as real signals of harm. Our experiences indicate that engaging in affective attunement can shield against affect abstraction and resulting alienation. Because the other components in our framework have bidirectional relationships with affect abstraction, it is an ideal intervention point. 

Since affect abstraction creates distance between researchers and the harms of the norms they follow, affective attunement can reduce that distance. By centering researchers' emotional discomfort with research practices, it makes space to question whether those practices are necessary, challenge their assumptions and values, and identify alternatives. We found that our experiences of alienation, particularly through epistemic injustice, were most acute when we failed to acknowledge our emotional responses. In these moments, we were subjecting ourselves to epistemic coercion; by succumbing to pressure to act in ways that contradicted our perceptions, we came to treat dominant interpretations of our circumstances as more legitimate than our own~\cite{Dandelet2021EpistemicCoercion}. By rendering harms visible, affective attunement expands which patterns of justification are ultimately considered reasonable. This makes it a powerful form of resisting ungrounded abstraction. Affective attunement complements other interventions that seek to amplify researchers' lived experiences in CS research spaces through storytelling~\citep{ogbonnaya2020critical, ymous2020terrified}, centering the knowledge of those who experience harm~\citep{birhane2021algorithmic, keyes2018misgendering, scheuerman2018safe}, and reflecting on one's own positionality and accountability~\citep{klumbyte2022critical}.

We recognize that centering affective attunement risks emphasizing individual correction over systemic accountability. This may shift responsibility onto marginalized people or those with less institutional power, burdening them with additional emotional and interpretive labor. We emphasize that affective attunement should operate collectively, such as in a research lab, where shared practices of collectively identifying and responding to emotional discomfort can establish solidarity and collective accountability.

\paragraph{Situating Our Work As Intervention.}
We see this work as an intervention to resist alienation that occurs from ungrounded abstraction. For us, the conversations we shared during analysis and consensus-building have helped build a community with intellectual safety. We adopted a methodology that explicitly resists epistemic injustice by claiming evidentiary value in our lived experiences. As the vignettes illustrate, academic papers by critical scholars also helped create the conditions for us to reflect on the academic tensions we experienced. Particularly critical for us was being able to name the structural factors driving these experiences. In the hope that we can do the same for others, we offer our interpretive framework as a hermeneutic resource for making meaning of one's own experiences of alienation.

\section{Positionality Statement}
The meaning we construct in this work emerges as we occupy roles where we often feel like the only people with our particular lived experiences and perspectives within our research communities. We recognize that these positions create a mix of hyper-visibility and isolation that shapes how we perceive others and how others perceive us. Among the authors, all three identify as women. Two are Indian. One author is mixed-race and Hispanic. One is queer. One is disabled. We have all inhabited elite research institutions that pride themselves on being global leaders in AI innovation while remaining deeply steeped in and sustained by AI extractivism. Our work is therefore shaped by shared academic training rooted in institutional contradiction, where we learn from and gain power from the very systems we seek to resist. These experiences collectively shape our attentiveness to power and exclusion within the spaces we inhabit and intrinsically influence the findings from our qualitative method.

\section{Ethical Considerations and Adverse Impact Statement}
This work draws on our real experiences with interacting with various individuals and communities. We recognize that sharing these experiences may raise ethical concerns, including risks to privacy or the possibility of presenting accounts without the perspectives of those involved. To mitigate these risks, we deliberately avoid attributing experiences to identifiable individuals --- both ourselves and the people we discuss. Instead, we use generalized descriptions, underspecified language, and a collective narrative voice to minimize the risk that any specific person or community can be identified. 

This deliberate underspecification creates a tradeoff with interpretive validity. Many contextual factors that could materially shape experiences of alienation, such as lab type, disciplinary subfield, or the geographic and political context of events referenced in our accounts, are intentionally omitted. The specification of the authors' universities provides limited institutional and disciplinary grounding for our interpretive claims, as a safer contextual proxy. We acknowledge that this constraint limits the degree to which readers can evaluate or situate particular accounts.

This project does not involve human subjects research as defined in 45 CFR 46.102(l) and therefore does not require review by an IRB.

\section{Acknowledgments}
We are extremely grateful to the anonymous reviewers and the following individuals for their feedback at various stages of this project: Haley Lepp, Angelina Wang, Sherri Rose, Dan Jurafsky, Jamie Lu, Meera Desai, Evan Dong, Nicky Kriplani, Thalia Zhang, Agata Foryciarz, Zander Majercik, Maria Luiza Rocha Bueno, and Britney Tran. We further appreciate the opportunity to present our work to and receive feedback from the members of the AI, Policy, and Practice Initiative (AIPP) at Cornell University and members of the Health Policy Data Science Lab at Stanford University. Finally, we are indebted to the many people --- peers, mentors, and loved ones --- who shaped our thinking on these topics over years of conversation. 

\bibliography{references}

@article{Colburn2007Abstraction,
  title        = {Abstraction in Computer Science},
  author       = {Colburn, Timothy and Shute, Gary},
  journal      = {Minds and Machines},
  volume       = {17},
  number       = {2},
  pages        = {169--184},
  year         = {2007},
  publisher    = {Springer},
  doi          = {10.1007/s11023-007-9061-7}
}

@article{Dandelet2021EpistemicCoercion,
  title        = {Epistemic Coercion},
  author       = {Sophia Dandelet},
  journal      = {Ethics},
  volume       = {131},
  number       = {3},
  pages        = {489--510},
  year         = {2021},
  publisher    = {University of Chicago Press},
  doi          = {10.1086/713146}
}

@article{Berkes2024SlowConvergence,
  author  = {Berkes, Enrico and Marion, Monica M. and Milojević, Staša and Weinberg, Bruce A.},
  title   = {Slow convergence: Career impediments to interdisciplinary biomedical research},
  journal = {Proceedings of the National Academy of Sciences of the United States of America},
  year    = {2024},
  volume  = {121},
  number  = {32},
  pages   = {e2402646121},
  doi     = {10.1073/pnas.2402646121},
  pmid    = {39074264},
  pmcid   = {PMC11317606}
}

@misc{zheng2025interdisciplinaryphdsfacebarriers,
      title={Interdisciplinary PhDs face barriers to top university placement within their disciplines}, 
      author={Xiang Zheng and Anli Peng and Xi Hong and Cassidy R. Sugimoto and Chaoqun Ni},
      year={2025},
      eprint={2503.21912},
      archivePrefix={arXiv},
      primaryClass={cs.CY},
      url={https://arxiv.org/abs/2503.21912}, 
}

@article{Makinen2025Interdisciplinary,
  author  = {M{\r a}kinen, Elina I. and Evans, Eliza D. and McFarland, Daniel A.},
  title   = {Interdisciplinary Research, Tenure Review, and Guardians of the Disciplinary Order},
  journal = {The Journal of Higher Education},
  year    = {2025},
  volume  = {96},
  number  = {1},
  pages   = {54--81},
  doi     = {10.1080/00221546.2024.2301912}
}

@incollection{agre1997critical,
  author       = {Agre, Philip E.},
  title        = {Toward a Critical Technical Practice: Lessons Learned in Trying to Reform AI},
  booktitle    = {Social Science, Technical Systems, and Cooperative Work: Beyond the Great Divide},
  editor       = {Bowker, Geoffrey C. and Star, Susan Leigh and Gasser, Les and Turner, William},
  year         = {1997},
  publisher    = {Lawrence Erlbaum Associates},
  address      = {Mahwah, NJ, USA},
  isbn         = {978-0-8058-2403-2},
}

@inproceedings{10.1145/3600211.3604673,
author = {Shelby, Renee and Rismani, Shalaleh and Henne, Kathryn and Moon, AJung and Rostamzadeh, Negar and Nicholas, Paul and Yilla-Akbari, N'Mah and Gallegos, Jess and Smart, Andrew and Garcia, Emilio and Virk, Gurleen},
title = {Sociotechnical Harms of Algorithmic Systems: Scoping a Taxonomy for Harm Reduction},
year = {2023},
isbn = {9798400702310},
publisher = {Association for Computing Machinery},
address = {New York, NY, USA},
url = {https://doi.org/10.1145/3600211.3604673},
doi = {10.1145/3600211.3604673},
booktitle = {Proceedings of the 2023 AAAI/ACM Conference on AI, Ethics, and Society},
pages = {723–741},
numpages = {19},
location = {Montr\'{e}al, QC, Canada},
series = {AIES '23}
}

@book{Fricker2007EpistemicInjustice,
  title     = {Epistemic Injustice: Power and the Ethics of Knowing},
  author    = {Fricker, Miranda},
  year      = {2007},
  publisher = {Oxford University Press},
  address   = {Oxford}
}

@article{Dotson2011TrackingEpistemicViolence,
  title   = {Tracking Epistemic Violence, Tracking Practices of Silencing},
  author  = {Dotson, Kristie},
  journal = {Hypatia},
  volume  = {26},
  number  = {2},
  pages   = {236--257},
  year    = {2011}
}

@article{Williams2019Metaeugenics,
  author       = {Williams, Rua Mae},
  title        = {Metaeugenics and Metaresistance: From Manufacturing the ‘Includeable Body’ to Walking Away from the Broom Closet},
  journal      = {Canadian Journal of Children’s Rights / Revue Canadienne des Droits des Enfants},
  volume       = {6},
  number       = {1},
  pages        = {60--77},
  year         = {2019},
  doi          = {10.22215/cjcr.v6i1.1976},
  url          = {https://ojs.library.carleton.ca/index.php/cjcr/article/view/1976},
}

@incollection{EllisBochner2000,
  author    = {Ellis, Carolyn and Bochner, Arthur P.},
  title     = {Autoethnography, Personal Narrative, Reflexivity: Researcher as Subject},
  booktitle = {Handbook of Qualitative Research},
  editor    = {Denzin, Norman K. and Lincoln, Yvonna S.},
  edition   = {2},
  publisher = {Sage},
  address   = {Thousand Oaks, CA},
  year      = {2000},
  pages     = {733--768}
}

@article{ellis2011autoethnography,
  title={Autoethnography: An Overview},
  author={Ellis, Carolyn and Adams, Tony E and Bochner, Arthur P},
  journal={Historical Social Research},
  pages={273--290},
  year={2011},
  publisher={JSTOR}
}

@article{Anderson2006,
  author  = {Anderson, Leon},
  title   = {Analytic Autoethnography},
  journal = {Journal of Contemporary Ethnography},
  volume  = {35},
  number  = {4},
  pages   = {373--395},
  year    = {2006},
  doi     = {10.1177/0891241605280449}
}

@article{StahlkeWall2016,
  author  = {Stahlke Wall, Sarah},
  title   = {Toward a Moderate Autoethnography},
  journal = {Qualitative Research in Organizations and Management: An International Journal},
  volume  = {11},
  number  = {1},
  pages   = {6--19},
  year    = {2016},
  doi     = {10.1108/QROM-03-2015-1276}
}

@book{DastonGalison2007,
  author    = {Daston, Lorraine and Galison, Peter},
  title     = {Objectivity},
  publisher = {Zone Books},
  address   = {New York},
  year      = {2007}
}

@article{haraway1988situated,
 author = {Donna Haraway},
 journal = {Feminist Studies},
 number = {3},
 pages = {575--599},
 publisher = {Feminist Studies, Inc.},
 title = {Situated Knowledges: The Science Question in Feminism and the Privilege of Partial Perspective},
 volume = {14},
 year = {1988}
}

@article{anderson1995feminist,
 author = {Elizabeth Anderson},
 journal = {Hypatia},
 number = {3},
 pages = {50--84},
 publisher = {[Hypatia, Inc., Wiley]},
 title = {Feminist Epistemology: An Interpretation and a Defense},
 urldate = {2024-03-17},
 volume = {10},
 year = {1995}
}

@article{harding1992rethinking,
 author = {Sandra Harding},
 journal = {The Centennial Review},
 number = {3},
 pages = {437--470},
 publisher = {Michigan State University Press},
 title = {Rethinking Standpoint Epistemology: What is ``Strong Objectivity?"},
 volume = {36},
 year = {1992}
}

@article{harding1992after,
 author = {Sandra Harding},
 journal = {Social Research},
 number = {3},
 pages = {567--587},
 publisher = {The Johns Hopkins University Press},
 title = {After the Neutrality Ideal: Science, Politics, and ``Strong Objectivity"},
 volume = {59},
 year = {1992}
}

@article{collins1989social,
 author = {Patricia Hill Collins},
 journal = {Signs},
 number = {4},
 pages = {745--773},
 publisher = {University of Chicago Press},
 title = {The Social Construction of Black Feminist Thought},
 volume = {14},
 year = {1989}
}

@article{LugonesSpelman1983,
  author  = {Lugones, María and Spelman, Elizabeth V.},
  title   = {Have We Got a Theory for You! Feminist Theory, Cultural Imperialism and the Demand for ``The Woman’s Voice'''},
  journal = {Women’s Studies International Forum},
  volume  = {6},
  number  = {6},
  pages   = {573--581},
  year    = {1983},
  doi     = {10.1016/0277-5395(83)90019-5}
}

@book{Code1991,
  author    = {Code, Lorraine},
  title     = {What Can She Know? Feminist Theory and the Construction of Knowledge},
  publisher = {Cornell University Press},
  address   = {Ithaca, NY},
  year      = {1991}
}

@article{adam1993gendered,
  title={Gendered Knowledge—Epistemology and Artificial Intelligence},
  author={Adam, Alison},
  journal={{AI} \& Society},
  volume={7},
  pages={311--322},
  year={1993},
  publisher={Springer}
}

@article{adam2000deleting,
  title={Deleting the Subject: A Feminist Reading of Epistemology in Artificial Intelligence},
  author={Adam, Alison},
  journal={Minds and Machines},
  volume={10},
  number={2},
  pages={231--253},
  year={2000},
  publisher={Springer}
}

@incollection{Lorde1981,
  author    = {Lorde, Audre},
  title     = {The Uses of Anger: Women Responding to Racism},
  booktitle = {Sister Outsider: Essays and Speeches},
  publisher = {Crossing Press},
  address   = {Freedom, CA},
  year      = {1981},
}

@book{Nagel1986ViewFromNowhere,
  author    = {Nagel, Thomas},
  title     = {The View from Nowhere},
  year      = {1986},
  publisher = {Oxford University Press},
  address   = {New York}
}

@article{kalluri2025computer,
  title={Computer-vision research powers surveillance technology},
  author={Kalluri, Pratyusha Ria and Agnew, William and Cheng, Myra and Owens, Kentrell and Soldaini, Luca and Birhane, Abeba},
  journal={Nature},
  volume={643},
  number={8070},
  pages={73--79},
  year={2025},
  publisher={Nature Publishing Group UK London}
}

@article{star2007enacting,
  title={Enacting silence: Residual categories as a challenge for ethics, information systems, and communication},
  author={Star, Susan Leigh and Bowker, Geoffrey C},
  journal={Ethics and Information Technology},
  volume={9},
  number={4},
  pages={273--280},
  year={2007},
  publisher={Springer}
}

@article{gebru2024tescreal,
  title={The TESCREAL bundle: Eugenics and the promise of utopia through artificial general intelligence},
  author={Gebru, Timnit and Torres, {\'E}mile P},
  journal={First Monday},
  year={2024}
}

@article{malazita2019infrastructures,
  title={Infrastructures of abstraction: How computer science education produces anti-political subjects},
  author={Malazita, James W and Resetar, Korryn},
  journal={Digital Creativity},
  volume={30},
  number={4},
  pages={300--312},
  year={2019},
  publisher={Taylor \& Francis}
}

@inproceedings{selbst2019fairness,
  title={Fairness and Abstraction in Sociotechnical Systems},
  author={Selbst, Andrew D and Boyd, Danah and Friedler, Sorelle A and Venkatasubramanian, Suresh and Vertesi, Janet},
  booktitle={Proceedings of the Conference on Fairness, Accountability, and Transparency},
  pages={59--68},
  year={2019}
}

@article{peterson2023abstracted,
  title={Abstracted power and responsibility in computer science ethics education},
  author={Peterson, Tina L and Ferreira, Rodrigo and Vardi, Moshe Y},
  journal={IEEE Transactions on Technology and Society},
  volume={4},
  number={1},
  pages={96--102},
  year={2023},
  publisher={IEEE}
}

@article{longino2020science,
  title={Science as social knowledge: Values and objectivity in scientific inquiry},
  author={Longino, Helen E},
  year={2020},
  publisher={Princeton university press}
}

@article{douglas2000inductive,
  title={Inductive risk and values in science},
  author={Douglas, Heather},
  journal={Philosophy of science},
  volume={67},
  number={4},
  pages={559--579},
  year={2000},
  publisher={Cambridge University Press}
}

@book{bowker2000sorting,
  title={Sorting things out: Classification and its consequences},
  author={Bowker, Geoffrey C and Star, Susan Leigh},
  year={2000},
  publisher={MIT press}
}

@article{boggs1977revolutionary,
  title={Revolutionary process, political strategy, and the dilemma of power},
  author={Boggs, Carl},
  journal={Theory and Society},
  volume={4},
  number={3},
  pages={359--393},
  year={1977},
  publisher={JSTOR}
}

@incollection{Marx1844EstrangedLabourTucker,
  author    = {Marx, Karl},
  title     = {Estranged Labour},
  booktitle = {The Marx--Engels Reader},
  editor    = {Tucker, Robert C.},
  edition   = {2},
  publisher = {W. W. Norton},
  address   = {New York},
  pages     = {66--125},
  year      = {1978},
  note      = {Originally written in 1844}
}

@book{freire1970pedagogy,
  title="Pedagogy of the Oppressed",
  author="Freire, Paulo",
  year=1970,
  publisher="Continuum",
  address="New York"
}

@article{noel2023collective,
author = {Tiffany Karalis Noel and Aiko Minematsu and Nikki Bosca},
title ={Collective Autoethnography as a Transformative Narrative Methodology},
journal = {International Journal of Qualitative Methods},
volume = {22},
number = {},
pages = {16094069231203944},
year = {2023},
doi = {10.1177/16094069231203944},
URL = {https://doi.org/10.1177/16094069231203944},
eprint = {https://doi.org/10.1177/16094069231203944}
}

@book{chang2016collaborative,
  title={Collaborative autoethnography},
  author={Chang, Heewon and Ngunjiri, Faith and Hernandez, Kathy-Ann C},
  year={2016},
  publisher={Routledge}
}

@article{ngunjiri2010living,
  title={Living autoethnography: Connecting life and research},
  author={Ngunjiri, Faith Wambura and Hernandez, Kathy-Ann C and Chang, Heewon},
  journal={Journal of research practice},
  volume={6},
  number={1},
  pages={E1--E1},
  year={2010}
}

@inproceedings{bala2023towards,
  title={Towards critical heritage in the wild: Analysing discomfort through collaborative autoethnography},
  author={Bala, Paulo and Sanches, Pedro and Ces{\'a}rio, Vanessa and Le{\~a}o, Sarah and Rodrigues, Catarina and Nunes, Nuno Jardim and Nisi, Valentina},
  booktitle={Proceedings of the 2023 CHI Conference on Human Factors in Computing Systems},
  pages={1--19},
  year={2023}
}

@inproceedings{kaltenhauser2024playing,
  title={Playing with perspectives and unveiling the autoethnographic kaleidoscope in HCI--A literature review of autoethnographies},
  author={Kaltenhauser, Annika and Stefanidi, Evropi and Sch{\"o}ning, Johannes},
  booktitle={Proceedings of the 2024 CHI Conference on Human Factors in Computing Systems},
  pages={1--20},
  year={2024}
}

@article{nel2018relational,
  title={Relational collaborative autoethnography: Post-doctoral fellowship in South Africa},
  author={Nel, NM},
  journal={Participatory Educational Research},
  volume={5},
  number={1},
  pages={60--73},
  year={2018},
  publisher={{\"O}zgen KORKMAZ}
}

@article{rutter2023its,
author = {Nikki Rutter and Ecem Hasan and Anna Pilson and Emma Yeo},
title ={``It's the End of the PhD as We Know it, and We Feel Fine…Because Everything Is Fucked Anyway'': Utilizing Feminist Collaborative Autoethnography to Navigate Global Crises},
journal = {International Journal of Qualitative Methods},
volume = {22},
number = {},
pages = {16094069211019595},
year = {2023},
doi = {10.1177/16094069211019595},
URL = {https://doi.org/10.1177/16094069211019595},
eprint = {https://doi.org/10.1177/16094069211019595}
}

@article{kafar2014autoethnography,
  title={Autoethnography, storytelling, and life as lived: A conversation between Marcin Kafar and Carolyn Ellis},
  author={Kafar, Marcin and Ellis, Carolyn},
  journal={Przeglad Socjologii Jakosciowej},
  volume={10},
  number={3},
  pages={124--143},
  year={2014}
}

@article{bundy2023all,
author = {Jessica Bundy and Vanessa Rhodes and Mariah Brooks and Maria Brisbane and Natalie Delia Deckard},
title ={All Things Considered: A Collaborative Critical Autoethnography of Emerging Racialized Scholars},
journal = {International Journal of Qualitative Methods},
volume = {22},
number = {},
pages = {16094069231180168},
year = {2023},
doi = {10.1177/16094069231180168},
URL = {https://doi.org/10.1177/16094069231180168},
eprint = {https://doi.org/10.1177/16094069231180168}
}

@incollection{clough2007introduction,
  author    = {Clough, Patricia Ticineto},
  title     = {Introduction},
  booktitle = {The Affective Turn: Theorizing the Social},
  editor    = {Clough, Patricia Ticineto and Halley, Jean},
  year      = {2007},
  publisher = {Duke University Press},
  address   = {Durham, NC},
  pages     = {1--33}
}

@article{suchman2025ai,
  author  = {Suchman, Lucy and Gururaja, Sireesh and Widder, David Gray},
  title   = {AI interdisciplinarity as critical technical practice},
  journal = {Cambridge Forum on AI: Culture and Society},
  year    = {2025},
  volume  = {1},
  pages   = {e2},
  doi     = {10.1017/cfc.2025.10004},
  publisher = {Cambridge University Press}
}

@inproceedings{khan2025whole,
  author    = {Khan, R. and Virguez, Lilianny and Paccotacya-Yanque, Rosa and Mekhael, Thomas and Munoriyarwa, Allen and Salgado, Leslie and Basu, Debarati and Mapaling, Curwyn and Perez, Natalie and Gaudet, Yves and Larrondo, Paula},
  title     = {Whole-Person Education for AI Engineers},
  booktitle = {Proceedings of the Canadian Engineering Education Association (CEEA-ACEG) Conference},
  year      = {2025},
  address   = {Montreal, Canada},
  month     = jun,
  url       = {https://ojs.library.queensu.ca/index.php/PCEEA/article/view/19600}
}

@inproceedings{russo2024bridging,
author = {Russo, Mayra and Jorgensen, Mackenzie and Scott, Kristen M. and Xu, Wendy and Nguyen, Di H. and Finocchiaro, Jessie and Olckers, Matthew},
title = {Bridging Research and Practice Through Conversation: Reflecting on Our Experience},
year = {2024},
isbn = {9798400712227},
publisher = {Association for Computing Machinery},
address = {New York, NY, USA},
url = {https://doi.org/10.1145/3689904.3694705},
doi = {10.1145/3689904.3694705},
booktitle = {Proceedings of the 4th ACM Conference on Equity and Access in Algorithms, Mechanisms, and Optimization},
articleno = {7},
numpages = {11},
location = {San Luis Potosi, Mexico},
series = {EAAMO '24}
}

@inproceedings{hampton2021black,
author = {Hampton, Lelia Marie},
title = {Black Feminist Musings on Algorithmic Oppression},
year = {2021},
isbn = {9781450383097},
publisher = {Association for Computing Machinery},
address = {New York, NY, USA},
url = {https://doi.org/10.1145/3442188.3445929},
doi = {10.1145/3442188.3445929},
booktitle = {Proceedings of the 2021 ACM Conference on Fairness, Accountability, and Transparency},
pages = {1},
numpages = {1},
location = {Virtual Event, Canada},
series = {FAccT '21}
}

@article{keyes2019misgendering,
  author  = {Keyes, Os and Peil, Michael and Barlas, Pınar},
  title   = {The Misgendering Machines: Trans/HCI Implications of Automatic Gender Recognition},
  journal = {Proceedings of the ACM on Human-Computer Interaction},
  volume  = {3},
  number  = {CSCW},
  year    = {2019},
  pages   = {1--22},
  doi     = {10.1145/3359318}
}

@book{shew2023technoableism,
  author    = {Shew, Ashley},
  title     = {Against Technoableism: Rethinking Who Needs Improvement},
  year      = {2023},
  publisher = {W. W. Norton \& Company},
  address   = {New York, NY}
}

@inproceedings{hanna2020towards,
author = {Hanna, Alex and Denton, Remi and Smart, Andrew and Smith-Loud, Jamila},
title = {Towards a critical race methodology in algorithmic fairness},
year = {2020},
isbn = {9781450369367},
publisher = {Association for Computing Machinery},
address = {New York, NY, USA},
url = {https://doi.org/10.1145/3351095.3372826},
doi = {10.1145/3351095.3372826},
booktitle = {Proceedings of the 2020 Conference on Fairness, Accountability, and Transparency},
pages = {501–512},
numpages = {12},
location = {Barcelona, Spain},
series = {FAT* '20}
}

@article{mohamed2020decolonial,
  author  = {Mohamed, Shakir and Png, Marie-Therese and Isaac, William},
  title   = {Decolonial AI: Decolonial Theory as Sociotechnical Foresight in Artificial Intelligence},
  journal = {Philosophy \& Technology},
  year    = {2020},
  volume  = {33},
  pages   = {659--684},
  doi     = {10.1007/s13347-020-00405-8}
}

@online{beygelzimer2021introducing,
  author       = {Alina Beygelzimer and Yann Dauphin and Percy Liang and Jennifer Wortman Vaughan},
  title        = {Introducing the {NeurIPS} 2021 Paper Checklist},
  year         = {2021},
  month        = mar,
  day          = {26},
  organization = {Neural Information Processing Systems Conference},
  url          = {https://neuripsconf.medium.com/introducing-the-neurips-2021-paper-checklist-3220d6df500b},
}

@online{naacl2022_responsible_nlp_blog,
  author       = {{NAACL 2022 Organizing Committee}},
  title        = {New Checklist for Responsible NLP Research},
  date         = {2021-12-09},
  year = {2021},
  url          = {https://2022.naacl.org/blog/responsible-nlp-research-checklist/},
  organization = {NAACL 2022},
}

@misc{olteanu2023rai_impact_blog,
  author       = {Alexandra Olteanu and Michael Ekstrand and Carlos Castillo and Jina Suh},
  title        = {Responsible AI Research Needs Impact Statements Too},
  year = {2023},
  date = {2023-11-19},
  url          ={https://medium.com/@alexandra.olteanu/responsible-ai-research-needs-impact-statements-too-7b7141031faf},
  organization = {Medium},
}

@inproceedings{Schroeder2025Disclosure,
author = {Schroeder, Hope and Pareek, Akshansh and Barocas, Solon},
title = {Disclosure without Engagement: An Empirical Review of Positionality Statements at FAccT},
year = {2025},
isbn = {9798400714825},
publisher = {Association for Computing Machinery},
address = {New York, NY, USA},
url = {https://doi.org/10.1145/3715275.3732079},
doi = {10.1145/3715275.3732079},
booktitle = {Proceedings of the 2025 ACM Conference on Fairness, Accountability, and Transparency},
pages = {1195–1210},
numpages = {16},
location = {
},
series = {FAccT '25}
}

@inproceedings{Liu2022Examining,
author = {Liu, David and Nanayakkara, Priyanka and Sakha, Sarah Ariyan and Abuhamad, Grace and Blodgett, Su Lin and Diakopoulos, Nicholas and Hullman, Jessica R. and Eliassi-Rad, Tina},
title = {Examining Responsibility and Deliberation in AI Impact Statements and Ethics Reviews},
year = {2022},
isbn = {9781450392471},
publisher = {Association for Computing Machinery},
address = {New York, NY, USA},
url = {https://doi.org/10.1145/3514094.3534155},
doi = {10.1145/3514094.3534155},
booktitle = {Proceedings of the 2022 AAAI/ACM Conference on AI, Ethics, and Society},
pages = {424–435},
numpages = {12},
location = {Oxford, United Kingdom},
series = {AIES '22}
}

@article{liang2021embracing,
author = {Liang, Calvin A. and Munson, Sean A. and Kientz, Julie A.},
title = {Embracing Four Tensions in Human-Computer Interaction Research with Marginalized People},
year = {2021},
issue_date = {April 2021},
publisher = {Association for Computing Machinery},
address = {New York, NY, USA},
volume = {28},
number = {2},
issn = {1073-0516},
url = {https://doi.org/10.1145/3443686},
doi = {10.1145/3443686},
journal = {ACM Trans. Comput.-Hum. Interact.},
month = apr,
articleno = {14},
numpages = {47}
}

@inproceedings{Cambo2022Model,
author = {Cambo, Scott Allen and Gergle, Darren},
title = {Model Positionality and Computational Reflexivity: Promoting Reflexivity in Data Science},
year = {2022},
isbn = {9781450391573},
publisher = {Association for Computing Machinery},
address = {New York, NY, USA},
url = {https://doi.org/10.1145/3491102.3501998},
doi = {10.1145/3491102.3501998},
booktitle = {Proceedings of the 2022 CHI Conference on Human Factors in Computing Systems},
articleno = {572},
numpages = {19},
location = {New Orleans, LA, USA},
series = {CHI '22}
}

@article{bauer1990barriers,
  title={Barriers against interdisciplinarity: Implications for studies of science, technology, and society (STS},
  author={Bauer, Henry H},
  journal={Science, Technology, \& Human Values},
  volume={15},
  number={1},
  pages={105--119},
  year={1990},
  publisher={Sage Publications Sage CA: Thousand Oaks, CA}
}

@article{fahimi2024articulation,
author = {Fahimi, Miriam and Russo, Mayra and Scott, Kristen M. and Vidal, Maria-Esther and Berendt, Bettina and Kinder-Kurlanda, Katharina},
title = {Articulation Work and Tinkering for Fairness in Machine Learning},
year = {2024},
issue_date = {November 2024},
publisher = {Association for Computing Machinery},
address = {New York, NY, USA},
volume = {8},
number = {CSCW2},
url = {https://doi.org/10.1145/3686973},
doi = {10.1145/3686973},
journal = {Proc. ACM Hum.-Comput. Interact.},
month = nov,
articleno = {434},
numpages = {23}
}

@article{klein2025provocations,
  title={Provocations from the humanities for generative AI research},
  author={Klein, Lauren and Martin, Meredith and Brock, Andr{\'e} and Antoniak, Maria and Walsh, Melanie and Johnson, Jessica Marie and Tilton, Lauren and Mimno, David},
  journal={arXiv preprint arXiv:2502.19190},
  year={2025}
}

@inproceedings{klumbyte2022critical,
author = {Klumbyte, Goda and Draude, Claude and Taylor, Alex S.},
title = {Critical Tools for Machine Learning: Working with Intersectional Critical Concepts in Machine Learning Systems Design},
year = {2022},
isbn = {9781450393522},
publisher = {Association for Computing Machinery},
address = {New York, NY, USA},
url = {https://doi.org/10.1145/3531146.3533207},
doi = {10.1145/3531146.3533207},
booktitle = {Proceedings of the 2022 ACM Conference on Fairness, Accountability, and Transparency},
pages = {1528–1541},
numpages = {14},
location = {Seoul, Republic of Korea},
series = {FAccT '22}
}

@book{kuhn1977essential,
  title={The essential tension: Selected studies in scientific tradition and change},
  author={Kuhn, Thomas S},
  year={1977},
  publisher={University of Chicago Press}
}

@inproceedings{Singh2025Exploring,
author = {Singh, Aneesha and Dechant, Martin Johannes and Patel, Dilisha and Soubutts, Ewan and Barbareschi, Giulia and Ayobi, Amid and Newhouse, Nikki},
title = {Exploring Positionality in HCI: Perspectives, Trends, and Challenges},
year = {2025},
isbn = {9798400713941},
publisher = {Association for Computing Machinery},
address = {New York, NY, USA},
url = {https://doi.org/10.1145/3706598.3713280},
doi = {10.1145/3706598.3713280},
booktitle = {Proceedings of the 2025 CHI Conference on Human Factors in Computing Systems},
articleno = {451},
numpages = {18},
location = {
},
series = {CHI '25}
}

@misc{facct2026authorguide,
  title        = {ACM Conference on Fairness, Accountability, and Transparency (FAccT) 2026 Author Guide},
  author       = {{ACM FAccT}},
  year         = {2026},
  howpublished = {\url{https://facctconference.org/2026/authorguide.html}},
}

@inproceedings{Widder2024Epistemic,
author = {Widder, David Gray},
title = {Epistemic Power in AI Ethics Labor: Legitimizing Located Complaints},
year = {2024},
isbn = {9798400704505},
publisher = {Association for Computing Machinery},
address = {New York, NY, USA},
url = {https://doi.org/10.1145/3630106.3658973},
doi = {10.1145/3630106.3658973},
booktitle = {Proceedings of the 2024 ACM Conference on Fairness, Accountability, and Transparency},
pages = {1295–1304},
numpages = {10},
location = {Rio de Janeiro, Brazil},
series = {FAccT '24}
}

@inproceedings{raji2021you,
  title={You can't sit with us: exclusionary pedagogy in AI ethics education},
  author={Raji, Inioluwa Deborah and Scheuerman, Morgan Klaus and Amironesei, Razvan},
  booktitle={Proceedings of the 2021 ACM conference on fairness, accountability, and transparency},
  pages={515--525},
  year={2021}
}

@inproceedings{birhane2022forgotten,
  title={The forgotten margins of AI ethics},
  author={Birhane, Abeba and Ruane, Elayne and Laurent, Thomas and S. Brown, Matthew and Flowers, Johnathan and Ventresque, Anthony and L. Dancy, Christopher},
  booktitle={Proceedings of the 2022 ACM conference on fairness, accountability, and transparency},
  pages={948--958},
  year={2022}
}

@article{metcalf2019owning,
  title={Owning ethics: Corporate logics, silicon valley, and the institutionalization of ethics},
  author={Metcalf, Jacob and Moss, Emanuel and others},
  journal={Social Research: An International Quarterly},
  volume={86},
  number={2},
  pages={449--476},
  year={2019},
  publisher={Johns Hopkins University Press}
}

@article{green2021contestation,
  title={The contestation of tech ethics: A sociotechnical approach to technology ethics in practice},
  author={Green, Ben},
  journal={Journal of Social Computing},
  volume={2},
  number={3},
  pages={209--225},
  year={2021},
  publisher={TUP}
}

@inproceedings{bietti2020ethics,
  title={From ethics washing to ethics bashing: A view on tech ethics from within moral philosophy},
  author={Bietti, Elettra},
  booktitle={Proceedings of the 2020 conference on fairness, accountability, and transparency},
  pages={210--219},
  year={2020}
}

@book{Williams2025Disabling,
  author    = {Williams, Rua M.},
  title     = {Disabling Intelligences: Legacies of Eugenics and How We are Wrong about AI},
  year      = {2025},
  publisher = {Springer Nature Switzerland},
  address   = {Cham},
  doi       = {10.1007/978-3-032-02665-1},
  isbn      = {978-3-032-02665-1}
}

@book{Chan2025Predatory,
  author    = {Chan, Anita Say},
  title     = {Predatory Data: Eugenics in Big Tech and Our Fight for an Independent Future},
  year      = {2025},
  publisher = {University of California Press},
  address   = {Oakland, CA},
  isbn      = {9780520402843}
}

@inproceedings{Birhane2022Values,
author = {Birhane, Abeba and Kalluri, Pratyusha and Card, Dallas and Agnew, William and Dotan, Ravit and Bao, Michelle},
title = {The Values Encoded in Machine Learning Research},
year = {2022},
isbn = {9781450393522},
publisher = {Association for Computing Machinery},
address = {New York, NY, USA},
url = {https://doi.org/10.1145/3531146.3533083},
doi = {10.1145/3531146.3533083},
booktitle = {Proceedings of the 2022 ACM Conference on Fairness, Accountability, and Transparency},
pages = {173–184},
numpages = {12},
location = {Seoul, Republic of Korea},
series = {FAccT '22}
}

@inproceedings{Laufer2023Optimization,
author = {Laufer, Benjamin and Gilbert, Thomas and Nissenbaum, Helen},
title = {Optimization's Neglected Normative Commitments},
year = {2023},
isbn = {9798400701924},
publisher = {Association for Computing Machinery},
address = {New York, NY, USA},
url = {https://doi.org/10.1145/3593013.3593976},
doi = {10.1145/3593013.3593976},
booktitle = {Proceedings of the 2023 ACM Conference on Fairness, Accountability, and Transparency},
pages = {50–63},
numpages = {14},
location = {Chicago, IL, USA},
series = {FAccT '23}
}

@inproceedings{Wang2025Identities,
author = {Wang, Angelina},
title = {Identities are not Interchangeable: The Problem of Overgeneralization in Fair Machine Learning},
year = {2025},
isbn = {9798400714825},
publisher = {Association for Computing Machinery},
address = {New York, NY, USA},
url = {https://doi.org/10.1145/3715275.3732033},
doi = {10.1145/3715275.3732033},
booktitle = {Proceedings of the 2025 ACM Conference on Fairness, Accountability, and Transparency},
pages = {485–497},
numpages = {13},
location = {
},
series = {FAccT '25}
}

@misc{aies2026authorguide,
  title        = {9th AAAI Conference on AI, Ethics, and Society Call for Papers},
  author       = {{AAAI AIES}},
  year         = {2026},
  howpublished = {\url{https://www.aies-conference.com/2026/call-for-papers/}},
}

@inproceedings{acuna2021are,
author = {Acuna, Daniel E. and Liang, Lizhen},
title = {Are AI Ethics Conferences Different and More Diverse Compared to Traditional Computer Science Conferences?},
year = {2021},
isbn = {9781450384735},
publisher = {Association for Computing Machinery},
address = {New York, NY, USA},
url = {https://doi.org/10.1145/3461702.3462616},
doi = {10.1145/3461702.3462616},
booktitle = {Proceedings of the 2021 AAAI/ACM Conference on AI, Ethics, and Society},
pages = {307–315},
numpages = {9},
location = {Virtual Event, USA},
series = {AIES '21}
}

@book{ahmed2004cultural,
  title={The Cultural Politics of Emotion},
  author={Ahmed, S.},
  isbn={9780748618477},
  url={https://books.google.com/books?id=9eRgQgAACAAJ},
  year={2004},
  publisher={Edinburgh University Press}
}

@inproceedings{ogbonnaya2020critical,
author = {Ogbonnaya-Ogburu, Ihudiya Finda and Smith, Angela D.R. and To, Alexandra and Toyama, Kentaro},
title = {Critical Race Theory for HCI},
year = {2020},
isbn = {9781450367080},
publisher = {Association for Computing Machinery},
address = {New York, NY, USA},
url = {https://doi.org/10.1145/3313831.3376392},
doi = {10.1145/3313831.3376392},
booktitle = {Proceedings of the 2020 CHI Conference on Human Factors in Computing Systems},
pages = {1–16},
numpages = {16},
location = {Honolulu, HI, USA},
series = {CHI '20}
}

@article{birhane2021algorithmic,
  title={Algorithmic injustice: A relational ethics approach},
  author={Birhane, Abeba},
  journal={Patterns},
  volume={2},
  number={2},
  year={2021},
  publisher={Elsevier}
}

@inproceedings{ymous2020terrified,
author = {Ymous, Anon and Spiel, Katta and Keyes, Os and Williams, Rua M. and Good, Judith and Hornecker, Eva and Bennett, Cynthia L.},
title = {``I am just terrified of my future'' — {E}pistemic Violence in Disability Related Technology Research},
year = {2020},
isbn = {9781450368193},
publisher = {Association for Computing Machinery},
address = {New York, NY, USA},
url = {https://doi.org/10.1145/3334480.3381828},
doi = {10.1145/3334480.3381828},
booktitle = {Extended Abstracts of the 2020 CHI Conference on Human Factors in Computing Systems},
pages = {1–16},
numpages = {16},
location = {Honolulu, HI, USA},
series = {CHI EA '20}
}

@article{keyes2018misgendering,
author = {Keyes, Os},
title = {The Misgendering Machines: Trans/HCI Implications of Automatic Gender Recognition},
year = {2018},
issue_date = {November 2018},
publisher = {Association for Computing Machinery},
address = {New York, NY, USA},
volume = {2},
number = {CSCW},
url = {https://doi.org/10.1145/3274357},
doi = {10.1145/3274357},
journal = {Proc. ACM Hum.-Comput. Interact.},
month = nov,
articleno = {88},
numpages = {22}
}

@article{scheuerman2018safe,
author = {Scheuerman, Morgan Klaus and Branham, Stacy M. and Hamidi, Foad},
title = {Safe Spaces and Safe Places: Unpacking Technology-Mediated Experiences of Safety and Harm with Transgender People},
year = {2018},
issue_date = {November 2018},
publisher = {Association for Computing Machinery},
address = {New York, NY, USA},
volume = {2},
number = {CSCW},
url = {https://doi.org/10.1145/3274424},
doi = {10.1145/3274424},
journal = {Proc. ACM Hum.-Comput. Interact.},
month = nov,
articleno = {155},
numpages = {27}
}

@book{farmer2004pathologies,
  title={Pathologies of power: Health, human rights, and the new war on the poor},
  author={Farmer, Paul},
  volume={4},
  year={2004},
  publisher={Univ of California Press}
}

@article{galtung1969violence,
  title={Violence, peace, and peace research},
  author={Galtung, Johan},
  journal={Journal of peace research},
  volume={6},
  number={3},
  pages={167--191},
  year={1969},
  publisher={Sage Publications Sage CA: Thousand Oaks, CA}
}

@book{nixon2011slow,
  title={Slow Violence and the Environmentalism of the Poor},
  author={Nixon, Rob},
  year={2011},
  publisher={Harvard University Press}
}

@article{stark2019facial,
  title={Facial recognition is the plutonium of AI},
  author={Stark, Luke},
  journal={XRDS: Crossroads, The ACM Magazine for Students},
  volume={25},
  number={3},
  pages={50--55},
  year={2019},
  publisher={ACM New York, NY, USA}
}

@article{selinger2022amazon,
  title={Amazon's ring: Surveillance as a slippery slope service},
  author={Selinger, Evan and Durant, Darrin},
  journal={Science as culture},
  volume={31},
  number={1},
  pages={92--106},
  year={2022},
  publisher={Taylor \& Francis}
}

@article{waelen2024ethics,
  title={The ethics of computer vision: An overview in terms of power},
  author={Waelen, Rosalie A},
  journal={AI and Ethics},
  volume={4},
  number={2},
  pages={353--362},
  year={2024},
  publisher={Springer}
}

@article{suchman2015situational,
  title={Situational Awareness: Deadly Bioconvergence at the Boundaries of Bodies and Machines},
  author={Suchman, Lucy},
  journal={MediaTropes},
  volume={5},
  number={1},
  year={2015}
}

@book{leslie1993cold,
  title={The Cold War and American science: The military-industrial-academic complex at MIT and Stanford},
  author={Leslie, Stuart W},
  year={1993},
  publisher={Columbia University Press}
}

@incollection{scheperHughesBourgois2004MakingSense,
  author    = {Scheper-Hughes, Nancy and Bourgois, Philippe},
  title     = {Introduction: Making Sense of Violence},
  booktitle = {Violence in War and Peace: An Anthology},
  editor    = {Scheper-Hughes, Nancy and Bourgois, Philippe},
  pages     = {1--27},
  publisher = {Blackwell},
  address   = {Oxford},
  year      = {2004}
}

@article{galtung1990CulturalViolence,
  author  = {Galtung, Johan},
  title   = {Cultural Violence},
  journal = {Journal of Peace Research},
  volume  = {27},
  number  = {3},
  pages   = {291--305},
  year    = {1990},
  doi     = {10.1177/0022343390027003005}
}

@inproceedings{young2022confronting,
  title={Confronting power and corporate capture at the FAccT Conference},
  author={Young, Meg and Katell, Michael and Krafft, PM},
  booktitle={Proceedings of the 2022 ACM Conference on Fairness, Accountability, and Transparency},
  pages={1375--1386},
  year={2022}
}

@book{husserl1970crisis,
  author    = {Husserl, Edmund},
  title     = {The Crisis of European Sciences and Transcendental Phenomenology:
               An Introduction to Phenomenological Philosophy},
  translator = {Carr, David},
  year      = {1970},
  publisher = {Northwestern University Press},
  address   = {Evanston, IL},
  isbn      = {9780810104587},
}

@inproceedings{hofmann2020,
author = {Hofmann, Megan and Kasnitz, Devva and Mankoff, Jennifer and Bennett, Cynthia L},
title = {Living Disability Theory: Reflections on Access, Research, and Design},
year = {2020},
isbn = {9781450371032},
publisher = {Association for Computing Machinery},
address = {New York, NY, USA},
url = {https://doi.org/10.1145/3373625.3416996},
doi = {10.1145/3373625.3416996},
booktitle = {Proceedings of the 22nd International ACM SIGACCESS Conference on Computers and Accessibility},
articleno = {4},
numpages = {13},
location = {Virtual Event, Greece},
series = {ASSETS '20}
}

@incollection{spivak1988subaltern,
  author    = {Spivak, Gayatri Chakravorty},
  title     = {Can the Subaltern Speak?},
  booktitle = {Marxism and the Interpretation of Culture},
  editor    = {Nelson, Cary and Grossberg, Lawrence},
  publisher = {University of Illinois Press},
  address   = {Urbana},
  year      = {1988},
  pages     = {271--313}
}

@article{malik2022critical,
  title={Critical technical awakenings},
  author={Malik, Maya and Malik, Momin M},
  journal={Journal of Social Computing},
  volume={2},
  number={4},
  pages={365--384},
  year={2022},
  publisher={TUP}
}

\appendix

\section{Thematic Similarities and Differences Across Vignettes} \label{app:themes}

Through our coding of the vignettes, several thematic similarities emerged, which are summarized in Table~\ref{tab:appendix-alienation-themes}. These themes demonstrate a bridge between the vignettes and the framework, since each theme closely relates to one or multiple components in the framework. While each theme may relate to framework components beyond what we have listed here, we indicate the framework component(s) that we see to be most closely related to each theme. 

\begin{table*}[t]
\centering
\small
\setlength{\tabcolsep}{4pt}
\renewcommand{\arraystretch}{1.2}
\caption{Themes of alienation across the three vignettes, with the framework components that each theme most corresponds to. The vignettes in Sections~\ref{sec:v-clean}-\ref{sec:v-cv} illustrate instances of alienation, while the vignette in Section~\ref{sec:v-safety} contrasts these dynamics through an example of intellectual safety.}
\label{tab:appendix-alienation-themes}
\begin{tabularx}{\textwidth}{
  >{\raggedright\arraybackslash}p{0.23\textwidth}
  >{\raggedright\arraybackslash}X
  >{\raggedright\arraybackslash}X
  >{\raggedright\arraybackslash}X
  >{\raggedright\arraybackslash}X}
\toprule
\textbf{Alienation Theme} & \textbf{Vignette \S~\ref{sec:v-clean}} & \textbf{Vignette \S~\ref{sec:v-cv}} & \textbf{Vignette \S~\ref{sec:v-safety}} & \textbf{Corresponding Framework Component(s)} \\
\midrule

\textit{Critiquing underlying assumptions of research agendas is constrained.}
&
The assignment's grading rubric required students to drop ``noisy'' data points, precluding critique of this step. 
&
Critique threatens a line of work that has accumulated prestige and heavy investment.
&
Explicitly stating and critiquing assumptions was welcome in the discussion.
& 
\textbf{Purpose abstraction:} why is an action taken in research?

\textbf{Position abstraction:} who is affected?

\textbf{Testimony abstraction:} whose voices and knowledge count in research?  
\\

\textit{Dissent is suppressed by norms of legitimacy.}
&
The assignment structure, the speed of the academic term, and the power differential between professors and students make critique of the assignment difficult.
&
The norms that define ``prestigious'' research suppress critique.
&
Turn-taking and active listening create space for all participants, and critique is encouraged rather than penalized.
& 
\textbf{Legitimacy:} norms determining ``real'' AI research and researchers.

\textbf{Busyness:} norms of constant productivity and speed of output.
\\

\textit{Ethics is insufficiently operationalized.}
&
Ethical concerns are acknowledged but not meaningfully addressed in practice.
&
Ethics is treated as important in principle but kept abstract, and operationalized only through incremental change.
&
Ethics is the primary object of substantive discussion.
& 
\textbf{Purpose abstraction:} why is an action taken in research?
\\

\textit{Rhetoric is decoupled from practice.}
&
The assignment discusses ethical implications but does not pursue an implementation that addresses these. 
&
``AI for good'' narratives overlook the dual use of technologies in violent contexts.
&
Critical commitments are enacted through engagement with critical papers.
& 
\textbf{Purpose abstraction:} why is an action taken in research?
\\

\textit{Performance metrics displace the substantive goals of computational work.}
&
Model metrics and the leaderboard define success.
&
Objectives related to career advancement incentivize the pursuit of dominant research agendas.
&
Participation in the discussion is not tied to academic and professional evaluation.
& 
\textbf{Purpose abstraction:} why is an action taken in research?

\textbf{Busyness:} norms of constant productivity and speed of output.
\\

\textit{Fast pace of computational work limits critical reflection.}
&
The heavy workload of the course inhibits slow engagement and reflection on discomfort.
&
Productivity is prioritized over deep engagement with the real-world, downstream impacts of research.
&
The pace of discussion and sharing of speaking space allows for intentional articulation and contemplation of ideas.
& 
\textbf{Busyness:} norms of constant productivity and speed of output.
\\

\textit{Real-world harm is not centered in computational work.}
&
Data points are not seen as humans experiencing violence.
&
Military use is recognized but not treated as relevant to day-to-day research.
&
The harms of AI practices are named and interrogated.
& 
\textbf{Position abstraction:} who is affected?
\\

\textit{Context is stripped away from computational tasks.}
&
Individuals are reduced to de-contextualized data points.
&
Research becomes detached from the lived realities shaped by its practices and outputs.
&
Context is explicitly surfaced through interdisciplinary translation.
& 
\textbf{Position abstraction:} who is affected?

\textbf{Testimony abstraction:} whose voices and knowledge count in research?  
\\

\textit{Emotional discomfort is experienced individually and in isolation.}
&
Discomfort is individually internalized and suppressed to finish assignments.
&
We experienced dissonance between our emotional responses and research obligations that we did not perceive in many of our colleagues.
&
Emotional responses are openly shared and reflected on together afterward, enabling collective sense-making.
& 
\textbf{Affect abstraction:} how researchers feel and are expected to feel.
\\

\bottomrule
\end{tabularx}
\end{table*}

\section{Analysis of Framework Relationships} \label{app:relations}

We describe how the components of our framework relate to one another, forming reinforcing dynamics that perpetuate harmful preconditions, mechanisms of abstraction, and outcomes.

\paragraph{Epistemic Injustice $\leftrightarrow$ Legitimacy} The norm of legitimacy and the harm of epistemic injustice reinforce each other by defining who can meaningfully participate in computational AI research and what kinds of work are recognized as valid. Legitimacy norms specify which values, methods, and findings constitute knowledge, shaping whose testimony is considered. When critique does not align with dominant legitimacy expectations, critics are often reframed as insufficiently knowledgeable about computational AI research, constituting epistemic injustice. Over time, the repeated marginalization of certain knowers reinforces epistemic exclusion.

\paragraph{Sociotechnical Harms $\leftrightarrow$ Busyness}
The norm of busyness and sociotechnical harms shape one another in a reinforcing cycle, structuring research priorities. A culture that privileges fast-paced research favors approaches that are quick to implement. These approaches often come at the expense of contextual understanding, producing human cost when these AI systems miss critical aspects of their interaction with the world. Conversely, sociotechnical harms, especially those that are particularly acute, can intensify busyness norms by requiring an urgent solution that redirects resources away from a deliberative process to ground the AI technology in the larger social system.

\paragraph{Affect Abstraction $\leftrightarrow$ Metaeugenic Regime}
Affect abstraction is a mechanism of the metaeugenic regime and a condition for its reproduction. The regime implicitly selects for researchers who can tolerate fast-paced, ungrounded research and suppress moral discomfort, while perpetuating abstractions that encourage distance. Thus, the metaeugenic regime enables affect abstraction. Affect abstraction grants the regime's values of generality, purity, and productivity a sense of inevitability, making it difficult for researchers to classify harms as structural rather than incidental.

\paragraph{Affect Abstraction $\leftrightarrow$ Other Mechanisms of Abstraction}
Affect abstraction emerges from and enables other mechanisms of abstraction. Repeated testimony abstraction can lead researchers to internalize dismissal of their critique, causing them to discount their intellectual contributions and the emotional responses that initially motivated their critique. Purpose abstraction obscures the real-world stakes of work by emphasizing technological contributions over societal problems that the work claims to address. This enables position abstraction, reducing the people affected by AI systems to decontextualized variables, making it easier to ignore harm. In the other direction, affect abstraction enables these other mechanisms to continue unchallenged, by normalizing the harms they create.

\paragraph{Affect Abstraction $\leftrightarrow$ Harmful Outcomes}
Affect abstraction and the two harmful outcomes co-create each other. When researchers engage in abstraction that inflicts epistemic injustice or sociotechnical harm, they experience discomfort that can be minimized through further abstraction. The resulting emotional distancing helps them continue tolerating epistemic injustice, and to ignore how their practices contribute to oppression when the AI system impacts people. Given the bidirectional relationship that affect abstraction has with the preconditions, mechanisms, and outcomes, it emerges as a critical component for the reinforcing system.

\end{document}